\documentclass[fleqn,twoside,twocolumn,nofootinbib,showkeys]{revtex4} 
\usepackage[sec,nocpr]{ujp} 

\def\lsim{\
  \lower-1.2pt\vbox{\hbox{\rlap{$<$}\lower5pt\vbox{\hbox{$\sim$}}}}\ }
\def\gsim{\
  \lower-1.2pt\vbox{\hbox{\rlap{$>$}\lower5pt\vbox{\hbox{$\sim$}}}}\ }
\begin{document}
\title[Microstructure of He II in the presence of boundaries]
{MICROSTRUCTURE OF He II\\ IN THE PRESENCE OF BOUNDARIES}
\author{M.D. Tomchenko}
\affiliation{Bogolyubov Institute for Theoretical Physics, Nat. Acad. of Sci. of Ukraine}
\address{14b, Metrolohichna Str., Kyiv 03680, Ukraine}
\email{mtomchenko@bitp.kiev.ua}
\udk{538.941} \pacs{67.10.-j,  67.25.dj,\\[-3pt] 67.25.dt} \razd{\secvi}

\setcounter{page}{123}%

\autorcol{M.D.\hspace*{0.7mm}Tomchenko}%

\begin{abstract}
We have studied the microstructure of a system of interacting Bose
particles under zero boundary conditions and have found two possible
orderings. One ordering is traditional and is characterized by the
Bogolyubov dispersion law $E(k) \approx\sqrt{\left
(\!\frac{\hbar^{2} k^2}{2m}\!\right )^{\!2} +
  q n\nu_{3}(k)\frac{\hbar^2 k^2}{m}}$ ($q=1$) at a
weak interaction. The second one is new and is characterized by the
same dispersion law, but with $q=2^{-d}$, where $d$ is the number of
noncyclic coordinates. At a weak interaction, the ground-state
energy is less for the new solution. The boundaries affect the bulk
microstructure due to the difference of the topologies of closed and
open systems.
\end{abstract}
 \maketitle

\section{Introduction}

In the Nature, all systems are finite, and the formulas deal with
finite values. However, the explicit account for boundaries leads
frequently to significant difficulties in the analysis. Therefore,
the physicists use a special trick that allows one to avoid the
consideration of boundaries, namely the passage to the thermodynamic
limit (Т-limit): $N, V \rightarrow \infty$, $N/V={\rm const}$
\cite{balescu}. This is the passage to infinite $N$ and $ V$, rather
than to very large ones: we increase infinitely the size of the
system, the boundaries go to infinity and as if disappear. However,
this reasoning contains a simple error: while the system is
increasing,  the boundaries cannot disappear, since they represent
the topological property. By means of the continuous increase of a
system with boundaries, it is impossible to obtain an infinite
system without boundaries. The latter would occur at some
topological jump, which we cannot imagine. This means that the
passage to the T-limit \textit{assumes} that such jump does not
affect the bulk properties of the system. But it can affect in
principle (no general proof of the opposite exists). To clarify the
existence of such effect, it is necessary to study various systems.
The goal of the present work is to study whether the boundaries
affect the bulk microstructure of a Bose liquid and an interacting
Bose gas.

In three models of interacting bosons \cite{lieb1963,
 gaudin,stringari1996,zaremba1998,cazalilla2004}, the solutions for
a system with boundaries and a cyclic system are the same: for a
homogeneous system at a weak interaction, they coincide with the
Bogolyubov solution \cite{bog,bz}. In these models, the interatomic
interaction was considered point-like (see Section 7 for details).
But the interaction is not point-like in the Nature. We will find
the solutions for a non-point interaction of the general form with a
potential, which can be expanded in a Fourier series. It turns out
that two solutions exist: traditional and new. We will study both
solutions and compare them with the results of other models and
with experiment.

\section{Ground State of He II: Equations}

Several methods of determination of the wave function (WF) $\Psi_{0}$ of the ground state (GS) of
helium-II are available
\cite{bz,chester1967,camp1973,feenb,yuv1,gfmc1979,gfmc1981,krot1986,krotscheck1988,swf1994a,swf1994b,rovenchak2000,mt1}
(see also reviews \cite{rea,ceperley1995,revAJP}).
In order not to take the boundaries into account,
they used periodic boundary conditions  (BCs) with  passing to the T-limit. We will find
the solutions for a system in a box with zero BCs.  The
influence of boundaries  on the structure of the WF can be
nontrivial. Therefore, it is expedient to use one of the exact
analytic methods for the determination of $\Psi_{0}$
\cite{camp1973,feenb,yuv1,krot1986}, since the numerical methods
may not catch the effect. We will operate within the method of
collective variables \cite{yuv1,yuv2} used by us in
\cite{mt1,mt2}.

We transit to the collective variables
\begin{equation}\label{1-1}
\rho_{\textbf{k}} = \frac{1}{\sqrt{N}}
\sum\limits_{j=1}^{N}e^{-i\textbf{k}\textbf{r}_j} \quad
(\textbf{k}\not= 0),
\end{equation}
 where $N$ is the total number of atoms, and
$\textbf{r}_j$ are the coordinates of atoms. For convenience, we
introduce the following notation. The sums over the wave
vectors\vspace*{-3mm}
\begin{equation}\label{1-7}
\textbf{k}=2\pi \left (\!\frac{j_{x}}{L_{x}}, \frac{j_{y}}{L_{y}},
\frac{j_{z}}{L_{z}} \!\right )\!,
\end{equation}
where $j_{x}, j_{y}, j_{z}$ are integers, and  $L_{x}, L_{y}, L_{z}$
are the sizes of the system, are denoted by the index $(2\pi)$:
$\sum^{(2\pi)}$\!; whereas the sums\vspace*{-1mm}
\begin{equation}\label{1-11}
\textbf{k}=\pi \left (\!\frac{j_{x}}{L_{x}}, \frac{j_{y}}{L_{y}},
\frac{j_{z}}{L_{z}}\! \right )
\end{equation}
are denoted by the index $(\pi)$:  $\sum^{(\pi)}$.

Under the periodic BCs with the subsequent passing to the T-limit,
the solution for $\Psi_{0}$ was obtained by I. Yukhnovskii and I.
Vakarchuk \cite{yuv1}:\vspace*{-1mm}
\begin{equation}\label{1-2}
\Psi_{0} = A e^{S_{b}},
\end{equation}\vspace*{-9mm}
\[S_{b}= \sum\limits_{\textbf{k}_{1}\neq 0}^{(2\pi)}\frac{a_{2}
(\textbf{k}_{1})}{2!}\rho_{\textbf{k}_{1}}
\rho_{-\textbf{k}_{1}}+\]\vspace*{-7mm}
\[+ \sum\limits_{\textbf{k}_{1},\textbf{k}_{2}\neq
0}^{(2\pi)\textbf{k}_{1}+\textbf{k}_{2}\not= 0}
\frac{a_{3}(\textbf{k}_{1},\textbf{k}_{2})}{3!\sqrt{N}}
\rho_{\textbf{k}_{1}}\rho_{\textbf{k}_{2}}\rho_{-\textbf{k}_{1} -
\textbf{k}_{2}}+\]\vspace*{-7mm}
\[+\!\sum\limits_{\textbf{k}_{1},\textbf{k}_{2},\textbf{k}_{3}\neq
0}^{(2\pi)\!\!\textbf{k}_{1}+\textbf{k}_{2}+\textbf{k}_{3}\not= 0}
  \frac{a_{4}(\textbf{k}_{1},\textbf{k}_{2},\textbf{k}_{3})}{4!N}
 \rho_{\textbf{k}_1}\rho_{\textbf{k}_2}\rho_{\textbf{k}_3}
 \rho_{-\textbf{k}_1 - \textbf{k}_2 - \textbf{k}_3}+\]\vspace*{-8mm}
\begin{equation}\label{1-3}
+\, ... ,
\end{equation}
where the functions $a_{j}(\textbf{k}_1,... ,\textbf{k}_{j-1})$
satisfy the chain
 of equations\vspace*{-3mm}
\begin{equation}\label{1-4}
E^{b}_0 = \frac{N-1}{2N}n\nu_{3}(0) - \sum\limits_{\textbf{k}\not=
0}^{(2\pi)}\frac{n\nu_{3}(k)}{2N} - \sum\limits_{\textbf{k}\not=
0}^{(2\pi)}\frac{\hbar^{2}k^2}{2mN}a_{2}(\textbf{k}),
\end{equation}
\[\frac{n\nu_{3}(k)m}{\hbar^2} + a_{2}(\textbf{k})k^2   -
 a_{2}^{2}(\textbf{k})k^2 =\]\vspace*{-7mm}
\begin{equation}\label{1-5}
= \sum\limits_{\textbf{q}\not=
0}^{(2\pi)}\frac{a_{3}(\textbf{k},\textbf{q})}{N}
 (q^2+\textbf{k}\textbf{q}) +
  \sum\limits_{\textbf{q}\not= 0}^{(2\pi)} q^2
\frac{a_{4}(\textbf{q},-\textbf{q},\textbf{k})}{2N},
\end{equation}\vspace*{-5mm}
\begin{equation}\label{1-a3}
a_{3}(\textbf{k}, \textbf{q}) \approx
-2\frac{R(\textbf{k},\textbf{q})}{\epsilon_{0}(k)+\epsilon_{0}(q)+
\epsilon_{0}(\textbf{k}+\textbf{q})},
\end{equation}\vspace*{-7mm}
\[R(\textbf{k}, \textbf{q}) = \textbf{k}\textbf{q}a_{2}(\textbf{k})
 a_{2}(\textbf{q})-\textbf{k}(\textbf{k}+\textbf{q})a_{2}(\textbf{k})
 a_{2}(\textbf{k}+\textbf{q})\,-\]\vspace*{-9mm}
\begin{equation}\label{1-a3r}
-\,\textbf{q}(\textbf{k}+\textbf{q})a_{2}(\textbf{q})
 a_{2}(\textbf{k}+\textbf{q}),
\end{equation}\vspace*{-7mm}
\begin{equation}\label{1-e}
\epsilon_{0}(\textbf{k}) = k^{2}(1-2a_{2}(\textbf{k})).
\end{equation}
Here, $n=N/V$, $V$ is the volume of the system, $E_{0}^{b}$ is the ``bulk''
energy of the GS per atom, and
\begin{equation}\label{1-6}
\nu_{3}(\textbf{k}) \equiv \nu_{3}(k)
=\int\limits_{-L_{x}}^{L_{x}}\!\!dx
\int\limits_{-L_{y}}^{L_{y}}\!\!dy
\int\limits_{-L_{z}}^{L_{z}}\!\!dz
U_{3}(r)e^{-i\textbf{k}\textbf{r}}
\end{equation}
is the Fourier-transform of the interaction potential $U_{3}(r)$ of
two He$^4$ atoms. The equation for $a_{3}$ is written in the zero
approximation not involving the corrections $a_{4}$ and $a_{5}$, and
the equations for higher corrections $a_{j\geq 4}(\textbf{k}_1,...
,\textbf{k}_{j-1})$ are not presented.

Consider He II
in a rectangular vessel $L_{x}\times L_{y} \times L_{z}$ in size.
 The Hamiltonian  has the form
\begin{equation}\label{1-8}
\hat{H} = -\frac{\hbar^{2}}{2m}\sum\limits_{j}\triangle_{j}
+\frac{1}{2}\sum\limits_{ij}^{i \not= j}
U_{3}(|\textbf{r}_{i}-\textbf{r}_{j}|) + U_{2},
\end{equation}\vspace*{-7mm}
\[U_{2} = \sum\limits_{j}\left [U_{w}(x_{j}) + U_{w}(L_{x}-x_{j})
+U_{w}(y_{j})\,+ \right.\]\vspace*{-7mm}
\begin{equation}\label{1-9}
+ \left. U_{w}(L_{y}-y_{j}) + U_{w}(z_{j}) + U_{w}(L_{z}-z_{j})
\right ].
\end{equation}
Here, $U_{2}$ is the interaction potential of helium atoms with all
walls.  The potentials of the walls are similar by form
\cite{zar1977} to the interaction potential of two helium atoms. The
free surface of liquid helium creates an energy barrier also, and we
consider it as a wall. Below, we assume all barriers identical with
the simple potential
\begin{equation}\label{1-10}
 U_{w}(x) =
\left \{\!\! \begin{array}{lll}
U_{s}> 0 , & \   x\leq 0,   & \\[3mm]
0,  & \ x>0. &  \end{array} \right.
\end{equation}
We take $U_{s}$
finite, and the transition $U_{s}\rightarrow \infty$ will be made in
final formulas.

Atoms polarize walls, and, therefore, the ``ref\-lec\-ti\-ons'' of
atoms arise on the walls. But since the po\-ten\-ti\-al of a wall is
infinitely high, it absorbs the ref\-lec\-tions, and we omit them in
$\hat{H}$ (\ref{1-8}).

Works \cite{yuv1,yuv2} used the non-Hermitian
Bo\-go\-lyu\-bov--Zubarev Hamiltonian. Since the absence of the
Hermitian property induces the questions \cite{lev2011}, we will
start from the ordinary Hermitian Hamiltonian (\ref{1-8}).

Under zero BCs, the WF is nonzero inside the vessel and is equal to
zero outside of it and on the boun\-da\-ry. Therefore, we seek the
WF of helium atoms in\-si\-de the vessel. In this case, all
functions in the Schr\"{o}\-din\-ger equation should be expanded in
a Fourier series with regard for the fact that the coordinates
$\textbf{r}_{j}$ of atoms are defined only inside the vessel. Let
the system occupies a volume $x \in [0, L_{x}]$, $y \in [0, L_{y}]$,
$y \in [0, L_{z}]$. The function $F(\textbf{r}_{1},\textbf{r}_{2})
=\tilde{U}(|\textbf{r}_{1}-\textbf{r}_{2}|)$ can be expanded in a
Fourier series in three ways: as a function of independent arguments
$\textbf{r}_{1}$ and $\textbf{r}_{2}$,
 as a function of the argument $(\textbf{r}_{1}-\textbf{r}_{2})$ (modulus is considered as a part of the function),
or as a function of $|\textbf{r}_{1}-\textbf{r}_{2}|$. In the first case,
\begin{equation}\label{pot1}
\tilde{U}(|x_{1}-x_{2}|) \!=\! \frac{1}{L^{2}_{x}}
\sum\limits_{k_{j_{1}}k_{j_{2}}}^{(2\pi)}\nu_{2}(k_{j_{1}},k_{j_{2}})e^{ik_{j_{1}}x_{1}+ik_{j_{2}}x_{2}},
\end{equation}\vspace*{-7mm}
\begin{equation}\label{pot2}
\nu_{2}(k_{j_{1}},k_{j_{2}})\! =\! \int\limits_{0}^{L_{x}}dx_{1}
\int\limits_{0}^{L_{x}}dx_{2}
\tilde{U}(|x_{1}-x_{2}|)e^{-ik_{j_{1}}x_{1}-ik_{j_{2}}x_{2}}
\end{equation}
(for simplicity, we restrict ourselves by the one-dimensional case). After some transformations, we obtain
\begin{equation}\label{pot3}
\nu_{2}(k_{j_{1}},k_{j_{2}}) =
L_{x}\nu(k_{j_{1}})\delta_{k_{j_{1}},-k_{j_{2}}}+
\tilde{\nu}_{2}(k_{j_{1}},k_{j_{2}}),
\end{equation}\vspace*{-7mm}
\begin{equation}\label{pot4}
\nu(k_{j_{1}}) = \int\limits_{-L_{x}}^{L_{x}}dx
\tilde{U}(|x|)e^{-ik_{j_{1}}x}.
\end{equation}
For a cyclic system, we have
\begin{equation}\label{pot6}
\tilde{U}(|x_{1}-x_{2}|) \!=\!
U(|x_{1}-x_{2}|)\!+\!U(L_{x}-|x_{1}-x_{2}|)
\end{equation}
(on a ring, one atom acts on another one from two sides, which yields the ``image'' $U(L_{x}-|x_{1}-x_{2}|)$), and the
 nondiagonal part $\tilde{\nu}_{2}(k_{j_{1}},k_{j_{2}})$
is equal to zero. From whence (in 3D):
\begin{equation}\label{pot5}
\tilde{U}_{3}(|\textbf{r}_{1}-\textbf{r}_{2}|) = \frac{1}{V}
\sum\limits_{\textbf{k}}^{(2\pi)}\nu_{3}(\textbf{k})e^{i\textbf{k}(\textbf{r}_{1}-\textbf{r}_{2})}
\end{equation}
with $\nu_{3}(\textbf{k})$ (\ref{1-6}). Note that (\ref{1-6})
contains the potential \textit{without} images. At the transition to
the T-limit, the images in $\tilde{U}_{3}$ are omitted. Then
relations (\ref{1-6}) and (\ref{pot5}) set the standard
Fourier-transformation for the T-limit.

For the system with boundaries,
$\tilde{U}_{3}(|\textbf{r}_{1}-\textbf{r}_{2}|)=$
$=U_{3}(|\textbf{r}_{1}-\textbf{r}_{2}|)$ (no images) and
$\tilde{\nu}_{2}(k_{j_{1}},k_{j_{2}})\neq 0$. However, formulas
(\ref{1-6}) and (\ref{pot5}) are valid only for a potential with
images, and their application to a potential without images distorts
the initial potential: in the 1D case, series (\ref{pot5}) gives
potential (\ref{pot6}) with image instead of the initial potential
$U(|x_{1}-x_{2}|)$. This point is of importance, since the image
$U(L_{x}-|x_{1}-x_{2}|)$ is \textit{unphysical} for the system with
boundaries, and its consideration causes the \textit{closure} of the
system by interaction.

When the nondiagonal part $\tilde{\nu}_{2}(k_{j_{1}},k_{j_{2}})\neq
0$ for the system with boundaries is considered, then the Fourier
series (\ref{pot1}), (\ref{pot3}) reproduces the potential
$U(|x_{1}-x_{2}|)$ exactly, without image. But the calculation of
$\tilde{\nu}_{2}(k_{j_{1}},k_{j_{2}})$ is difficult. We can use a
Fourier series, where the vector
$(|x_{1}-x_{2}|,|y_{1}-y_{2}|,|z_{1}-z_{2}|)$ is the argument. But
then the moduli will enter also in the exponent, and we cannot use
the method of collective variables. Therefore, the best way is to
use the expansion
\begin{equation}\label{1-12}
U_{3}(|\textbf{r}_{1}-\textbf{r}_{2}|) = \frac{1}{2^{d}V}
\sum\limits_{\textbf{k}}^{(\pi)}\nu_{3}(\textbf{k})e^{i\textbf{k}(\textbf{r}_{1}-\textbf{r}_{2})},
\end{equation}
with $\nu_{3}(\textbf{k})$ (\ref{1-6}). Here, the expansion argument
is $(\textbf{r}_{1}-\textbf{r}_{2})$, and
 $x_{i}$ and $x_{j}$  take values from
$[0, L_{x}]$, whereas  $x_{i}-x_{j}\in [-L_{x}, L_{x}].$  By the rules of
Fourier analysis, the function $U_{3}(|\textbf{r}_{i}-\textbf{r}_{j}|)$ is expanded in series
(\ref{1-6}), (\ref{1-12}): with $\textbf{k}$ (\ref{1-11}) and with the volume equal to $2L_{x}\times 2L_{y}\times 2L_{z}=2^{d}V$.
This series is the simplest one, reproduces the initial potential \textit{exactly} (without images, in particular),
and is convenient for the method of collective variables.
For the total interatomic potential, we obtain
\[\frac{1}{2}\sum\limits_{i \not= j} U_{3}(|\textbf{r}_{i}-\textbf{r}_{j}|) =
 \sum\limits_{\textbf{k} \not= 0}^{(\pi)}\frac{n\nu_{3}(k)}{2^{d+1}}\rho_{\textbf{k}}\rho_{-\textbf{k}}\,+\]\vspace*{-5mm}
\begin{equation}\label{1-13}
+ \, (N-1)\frac{n\nu_{3}(0)}{2^{d+1}} -
 \sum\limits_{\textbf{k}\not= 0}^{(\pi)}\frac{n\nu_{3}(k)}{2^{d+1}}.
\end{equation}
Under periodic BCs (in the T-limit or without it), we have
\cite{bz}\vspace*{-3mm}
\[\frac{1}{2}\sum\limits_{i \not= j}
U_{3}(|\textbf{r}_{i}-\textbf{r}_{j}|) =
 \sum\limits_{\textbf{k} \not= 0}^{(2\pi)}\frac{n\nu_{3}(k)}{2}\rho_{\textbf{k}}\rho_{-\textbf{k}}\,+\]\vspace*{-5mm}
\begin{equation}\label{1-13b}
+\, (N-1)\frac{n\nu_{3}(0)}{2} - \sum\limits_{\textbf{k}\not=
0}^{(2\pi)}\frac{n\nu_{3}(k)}{2}
\end{equation}
(for a finite system, $U_{3}\rightarrow \tilde{U}_{3}$: we consider
the \mbox{images}).

This is a key point. If the traditional expansion (\ref{1-6}),
(\ref{pot5}), (\ref{1-13b}) is used, then we will obtain, obviously,
the traditional solution for $E_{0}$ and $E(k)$ (with small
``surface'' corrections). But if expansion (\ref{1-6}),
(\ref{1-12}), (\ref{1-13b}) is applied, we obtain, probably, a
solution with $\nu_{3}(k)\rightarrow \nu_{3}(k)/2^d$ and
$\sum^{(2\pi)}\rightarrow \sum^{(\pi)}$, since the bulk equations
(\ref{1-3})--(\ref{1-a3}) are ``generated'' by the total potential.
The solution cannot depend, of course, on the expansion. But some
solution is better seen with the use of certain variables, whereas
another one can be better seen in the language of other variables.
Since the traditional solution was studied many times, we will
consider a new solution following from expansion (\ref{1-6}),
(\ref{1-12}).

Let us determine the ground state WF, which satisfies the Schr\"{o}dinger equation
\begin{equation}\label{1-15}
\hat{H}\Psi_{0} = NE_{0}\Psi_{0}
\end{equation}
with Hamiltonian (\ref{1-8}), (\ref{1-13}). The main question is in
which form we should seek $\Psi_{0}.$ Solution (\ref{1-2}),
(\ref{1-3}) for periodic BCs is obtained from the requirement of the
translation invariance and the functional independence \cite{yuv1}
of the following collections of \mbox{variables
$\rho_{\textbf{k}}$:}
\begin{equation}\label{1-op}
\rho_{-\textbf{k}_{1}},~
\rho_{-\textbf{k}_{1}}\rho_{-\textbf{k}_{2}},~
\rho_{-\textbf{k}_{1}}\rho_{-\textbf{k}_{2}}\rho_{-\textbf{k}_{3}},~
 ...\,.
\end{equation}
But the boundaries break the translation invariance. However,
solution (\ref{1-2}), (\ref{1-3}) can be obtained without regard
for the translation invariance, by substituting the bare function
$\Psi_{0}=\rm const$, describing the free bosons, in the
Schr\"{o}dinger equation. The equation itself ``sets'' the form of
corrections, which appear  in the exponent. We will seek $\Psi_{0}$
analogously, by substituting the solution for free particles in the
Schr\"{o}dinger equation.

The WFs of the system of $N$ free Bose particles
located in a rectangular box with potential (\ref{1-9}), (\ref{1-10}) are known. For
identical states of all particles, the WFs are  \cite{land3}
\[\Psi^{f}_{\textbf{l}}=\prod\limits_{j=1}^{N}\left
[\sin{(k_{l_{x}}x_{j}+\delta_{l_{x}})}
\sin{(k_{l_{y}}y_{j}+\delta_{l_{y}})}\right. \times\]\vspace*{-7mm}
\begin{equation}\label{1-16}
\times  \left. \sin{(k_{l_{z}}z_{j}+\delta_{l_{z}})}\right ],
\end{equation}
where $l_{x}, l_{y}, l_{z}=1, 2, 3, ... ,$ and $k_{l_{x}}$ and
$\delta_{l_{x}}$ satisfy the equations
\[\sin{\delta_{l_{x}}}=\frac{\gamma_{x}k_{l_{x}}L_{x}}{2},\]
\begin{equation}\label{1-17}
\sin{(k_{l_{x}}L_{x}+\delta_{l_{x}})}=\pm
\frac{\gamma_{x}k_{l_{x}}L_{x}}{2},
\end{equation}
which yields
\begin{equation}\label{1-18}
k_{l_{x}}L_{x}=\pi l_{x} -
2\delta_{l_{x}},\quad\delta_{l_{x}}=\arcsin{(\gamma_{x}k_{l_{x}}L_{x}/2)}.
\end{equation}
Here, we denote
$\gamma_{x}=\frac{\hbar}{L_{x}}\sqrt{\frac{2}{mU_{s}}}$, and the
values of $\arcsin$ are taken between $0$ and $\pi/2$. For the
realistic systems, $\gamma_{x}\ll 1$. For example, for He$^4$ atoms
at $U_{s}=$ =~100\,K and $L_{x}=10$\,cm  we have $\gamma_{x}\approx
4.92\times 10^{-10}$.  At $l_{x}\ll 1/\pi\gamma_{x},$ relation
(\ref{1-18}) yields
\begin{equation}\label{1-20}
k_{l_{x}}\approx l_{x}k_{1x}=l_{x}(1-\gamma_{x})\pi /L_{x},
\end{equation}\vspace*{-9mm}
\begin{equation}\label{1-21}
\delta_{l_{x}}\approx l_{x}\delta_{1x}=l_{x}\pi \gamma_{x}/2.
\end{equation}
The maximally possible values of $\delta_{l_{x}}$ and $k_{l_{x}}$
are determined from the condition $\sin{\delta_{l_{x}}}=1$:
\begin{equation}\label{1-22}
\delta^{\rm max}_{l_{x}}=\pi/2,  \quad k^{\rm
max}_{l_{x}}=k_{1x}/\delta_{1x}.
\end{equation}
The WF of the GS of a single atom inside of the box,
\[\Psi^{f}_{0,1}(\textbf{r})\!=\!\sin(k_{1x}x\!+\!\delta_{1x})\sin(k_{1y}y\!+\!\delta_{1y})\sin(k_{1z}z\!+\!\delta_{1z}),\]
 is positive and nonzero. Outside of the box, the WF decays exponentially.

With regard for the above consideration, the WF of the ground state of $N$
interacting Bose particles located in the box can be sought
in the form
\begin{equation}\label{1-23}
\Psi_{0} = A_{1} \Psi^{f}_{\textbf{0}}e^{S_{w}^{(1)}+\tilde{S}_{b}},
\end{equation}
where $\Psi^{f}_{\textbf{0}}$ is the WF of the GS of $N$
\textit{free} Bose particles in the box (function (\ref{1-16}) with
$l_{x}=l_{y}=l_{z}=1$), $e^{\tilde{S}_{b}}$ is a WF of the form
(\ref{1-2}), (\ref{1-3}) with the sums  $\sum^{(\pi)}$ instead of
$\sum^{(2\pi)}$ (this is seen from expansion (\ref{1-13})), and
$e^{S_{w}^{(1)}}$ describes the cross terms arising at the
substitution of $\Psi^{f}_{\textbf{0}}e^{\tilde{S}_{b}}$ in the
Schr\"{o}dinger equation. By the theorem on nodes, if the GS is
nondegenerate, then $\Psi_{0}$ is nonzero in all points in the box.
Therefore, it is convenient to lift the function $S_{w}^{(1)}$ to
the exponent. The factor $e^{\tilde{S}_{b}}$ appears due to the
interatomic interaction, and $\Psi^{f}_{\textbf{0}}$  ensures the
fulfillment of BCs and the transition to the WF of free particles at
the switch-off of the interaction.

Note  that structure (\ref{1-23}) was proposed previously by
S.~Yushchenko \cite{yush}.

The function $\Psi_{0}$ (\ref{1-23}) must satisfy the Schr\"{o}dinger equation (\ref{1-15}) and be zero at the boundaries.
The last property is ensured  at $U_{s}\rightarrow \infty$
by the factor $\Psi^{f}_{\textbf{0}}$ which satisfies the equations
\begin{equation}\label{1-24}
 -\frac{\hbar^{2}}{2m}\sum\limits_{j}\triangle_{j}\Psi^{f}_{\textbf{0}} + U_{2}
 \Psi^{f}_{\textbf{0}} = NE_{0}^{f}\Psi^{f}_{\textbf{0}},
 \end{equation}\vspace*{-5mm}
 \begin{equation}\label{1-25}
 E_{0}^{f} = \frac{\hbar^{2}k_{1}^{2}}{2m}, \quad
 k_{1}^{2}=k_{1x}^{2}+k_{1y}^{2}+k_{1z}^{2}.
 \end{equation}
The  function $\tilde{S}_{b}$ satisfies the equation
\begin{equation}\label{1-26}
 -\frac{\hbar^{2}}{2m}\sum\limits_{j}\triangle_{j}e^{\tilde{S}_{b}} +
 \frac{1}{2}\sum\limits_{i \not= j} U_{3}(|\textbf{r}_{i}-\textbf{r}_{j}|)
 e^{\tilde{S}_{b}} = N\tilde{E}^{b}_{0} e^{\tilde{S}_{b}}.
 \end{equation}
 Then Eq. (\ref{1-15}) is reduced to the  equation for $S_{w}^{(1)}$:
\[ -\frac{\hbar^{2}}{2m}\sum\limits_{j}\left [ \triangle_{j}S_{w}^{(1)} +
 (\nabla_{j}S_{w}^{(1)})^{2}+
 2\nabla_{j}S_{w}^{(1)}\frac{\nabla_{j}\Psi^{f}_{\textbf{0}}}{\Psi^{f}_{\textbf{0}}}
 \,+\right. \]\vspace*{-5mm}
\[+ \left. 2\nabla_{j}\tilde{S}_{b}\left (\!\nabla_{j}S_{w}^{(1)}+
 \frac{\nabla_{j}\Psi^{f}_{\textbf{0}}}{\Psi^{f}_{\textbf{0}}}\!\right )\right
 ]=\]\vspace*{-5mm}
\begin{equation}\label{1-27}
= N(E_{0}-\tilde{E}_{0}^{b}-E_{0}^{f}),
\end{equation}\vspace*{-5mm}
\[ \frac{\nabla_{j}\Psi^{f}_{\textbf{0}}}{\Psi^{f}_{\textbf{0}}} =
 \textbf{i}_{x}k_{1x}\cot{(k_{1x}x_{j}+\delta_{1x})}\,+\]\vspace*{-5mm}
\begin{equation} \label{1-28}
 +\, \textbf{i}_{y}k_{1y}\cot(k_{1y}y_{j}\!+\!\delta_{1y})\! +\!
 \textbf{i}_{z}k_{1z}\cot(k_{1z}z_{j}\!+\!\delta_{1z}).
\end{equation}
By seeking $\Psi_{0}$ inside of the vessel, we set
$\cot{(k_{1x}x+\delta_{1x})}$ on the interval $[0, L_{x}]$. In
order to apply the method of collective variables, we expand the
cotangent in the Fourier series:
\begin{equation}\label{1-29}
 \cot{(k_{1x}x+\delta_{1x})} =
 \sum\limits_{j_{x}}^{(2\pi)}C_{1x}(j_{x})e^{i2\pi j_{x}x/L_{x}},
 \end{equation}
where $j_{x}$ runs all integers.
We have
\[C_{1x}(j_{x}) \!=\!  - C_{1x}(-j_{x})\! =\!
\int\limits_{0}^{L_{x}}\!
 \frac{dx}{L_{x}}\cot{(k_{1x}x\!+\!\delta_{1x})}e^{-iq_{x}x}=\]\vspace*{-7mm}
\begin{equation}\label{1-30}
 = -2i\int\limits_{0}^{1/2}
 dx\sin{(2\pi j_{x}x)}\cot{(\pi x -2\delta_{1x}x+\delta_{1x})}.
 \end{equation}
From whence, $C_{1x}(0)=0$, and the function $C_{1x}(j_{x})$ is
approximated by the formula
\begin{equation}
 C_{1x}(j_{x}>0) \approx -i + \frac{i4j_{x}\delta_{1x}}{\pi}
 \ln\left (\!1+ \frac{\pi}{4j_{x}\delta_{1x}}\!\right )\!,
      \label{1-31} \end{equation}
which is valid at $j_{x} \lsim j_{m} =  1/\delta_{1x}$ with an error
\mbox{of $\leq 3\%$.}

We note that $\cot{(k_{1x}x+\delta_{1x})}$ satisfies the
requirements to functions expanded in the Fourier series. At
$\delta_{1x}=0,$ we obtain $\cot{(\pi x/L_{x})}$, which does not
satisfy these requirements, since $\int_{0}^{L_{x}}|\cot{(\pi
x/L_{x})}|dx=$ $=\infty$. In view of this difficulty, we preserve
all $\delta$ to be nonzero in calculations and set $\delta
\rightarrow 0$ in final formulas.

Using the difference of the functional structure
 of the collections $\rho_{\textbf{k}}$ (\ref{1-op}),
we determine the structure of the solution for $S_{w}^{(1)}$ from (\ref{1-27})--(\ref{1-29}) and (\ref{1-3}):
\[S_{w}^{(1)}\!=\! \sum\limits_{\textbf{q}\neq 0}^{(2\pi)}S_{1}^{(1)}(\textbf{q})\rho_{-\textbf{q}} \!+\!
 \sum\limits_{\textbf{q},\textbf{q}_{1}\neq 0}^{ \textbf{q}+\textbf{q}_{1}\neq
 0}\!
\frac{S_{2}^{(1)}(\textbf{q},\textbf{q}_{1})}{\sqrt{N}}\rho_{\textbf{q}_{1}}\rho_{-\textbf{q}_{1}-\textbf{q}}\,+\]\vspace*{-7mm}
\begin{equation}\label{1-32}
+ \sum\limits_{\textbf{q},\textbf{q}_{1},\textbf{q}_{2}\neq 0}^{\textbf{q}+
 \textbf{q}_{1}+\textbf{q}_{2}\neq 0}
 \frac{S_{3}^{(1)}(\textbf{q},\textbf{q}_{1},\textbf{q}_{2})}{N}
 \rho_{\textbf{q}_{1}}\rho_{\textbf{q}_{2}}
 \rho_{-\textbf{q}_{1}-\textbf{q}_{2}-\textbf{q}}+...,
 \end{equation}
where $\textbf{q}$ and $\textbf{q}_{j}$ run values  (\ref{1-7}) and
(\ref{1-11}), respectively (the first fact follows from
(\ref{1-29}), and the second does from (\ref{1-12}), (49)).
 The functions $S_{j}^{(1)}$ satisfy the symmetry relations
\begin{equation}\label{1-32b}
 S_{2}^{(1)}(\textbf{q},\textbf{q}_{1}) = S_{2}^{(1)}(\textbf{q},-\textbf{q}-\textbf{q}_{1}),
 \end{equation}\vspace*{-7mm}
\[S_{3}^{(1)}(\textbf{q},\textbf{q}_{1},\textbf{q}_{2})= S_{3}^{(1)}(\textbf{q},\textbf{q}_{2},\textbf{q}_{1})=\]\vspace*{-7mm}
\[= S_{3}^{(1)}(\textbf{q},\textbf{q}_{1},-\textbf{q}-\textbf{q}_{1}-\textbf{q}_{2})=\]\vspace*{-7mm}
 \begin{equation}\label{1-32c}
=S_{3}^{(1)}(\textbf{q},-\textbf{q}-\textbf{q}_{1}-\textbf{q}_{2},\textbf{q}_{2}),
\end{equation}
and the analogous ones can be written for higher $S_{j}^{(1)}$.

The equations for  $S_{j}^{(1)}$ can be found from Eq. (\ref{1-27}),
if we substitute all functions in it and collect all terms referred
to each collection in (\ref{1-op}) and the constant. In view of the
functional difference of collections in (\ref{1-op}), we equate
these groups of terms to zero. This yields the chain of equations
for $E_{0}$  and the \mbox{functions $S_{j}^{(1)}$:}
\begin{equation}\label{1-33}
 E_{0}= \tilde{E}^{b}_0 + A_{1},
\end{equation}
\[A_{1}=  \frac{\hbar^2}{2m}\left [ k_{1}^{2} -
 \frac{1}{N}\sum\limits_{\textbf{q}\neq
 0}^{(2\pi)}q^{2}S_{1}^{(1)}(\textbf{q})S_{1}^{(1)}(-\textbf{q})\,
  -\right.\]\vspace*{-7mm}
\begin{equation}\label{1-33b}
\!-\!\left. \frac{i}{\sqrt{N}}\!\sum\limits_{q_{x}\neq
0}^{(2\pi)}\!2k_{1x}q_{x}C_{1x}(q_{x})S_{1}^{(1)}\!(-\textbf{q}_{x})
 \!+\! (x\!\rightarrow\! y, z)\!\right ]\!,
\end{equation}\vspace*{-7mm}
\[S_{1}^{(1)}(\textbf{q})\epsilon_{0}(\textbf{q})=\]\vspace*{-7mm}
\[= -i\sqrt{N} 2k_{1x}q_{x}a_{2}(-\textbf{q}_{x})C_{1x}(q_{x})\delta_{\textbf{q},\textbf{q}_{x}}
 + (x\rightarrow y, z)\,+\]\vspace*{-9mm}
\[+ \frac{2}{N}\sum\limits_{\textbf{q}_{1}\neq 0}^{(\pi)}\left \{(q_{1}^{2}+\textbf{q}_{1}\textbf{q})
 S_{2}^{(1)}(\textbf{q},\textbf{q}_{1})\,+ \right.\]\vspace*{-7mm}
\[+\,3q_{1}^{2}S_{3}^{(1)}(\textbf{q},\textbf{q}_{1},-\textbf{q}_{1})\,+\]\vspace*{-7mm}
\[
+\,\sqrt{N}2q_{1}^{2}S_{1}^{(1)}(\textbf{q}_{1})S_{2}^{(1)}(\textbf{q}-\textbf{q}_{1},\textbf{q}_{1})\,+\]\vspace*{-7mm}
\[+\left. 0.5\sqrt{N}(q_{1}^{2}-\textbf{q}_{1}\textbf{q})S_{1}^{(1)
}(\textbf{q}_{1})S_{1}^{(1)}(\textbf{q}-\textbf{q}_{1}) \right
\}+\]\vspace*{-7mm}
\[+ \Bigg \{\! \sum\limits_{p_{x}\neq 0}^{(2\pi)}2ik_{1x}C_{1x}(p_{x})
 \left [(q_{x}-p_{x})S_{1}^{(1)}(\textbf{q}-\textbf{p}_{x})\,- \right. \]\vspace*{-7mm}
 \begin{equation} \label{1-34}
-\left. 2p_{x}S_{2}^{(1)}(\textbf{q}-\textbf{p}_{x},-\textbf{q})
 \right ] + (x\rightarrow y, z) \!\Bigg \}\!,
\end{equation}\vspace*{-7mm}
\[S_{2}^{(1)}(\textbf{q},\textbf{q}_{1})\left [\epsilon_{0}(\textbf{q}_{1}) + \epsilon_{0}(\textbf{q}+\textbf{q}_{1})
\right ]+\]\vspace*{-7mm}
\[+\,2S_{1}^{(1)}(\textbf{q})a_{2}(-\textbf{q}_{1})\textbf{q}\textbf{q}_{1}-q^{2}S_{1}^{(1)}
(\textbf{q})a_{3}(\textbf{q},\textbf{q}_{1})=\]\vspace*{-7mm}
\[= \delta_{\textbf{q},\textbf{q}_{x}}\sqrt{N}ik_{1x}C_{1x}(q_{x})\left \{ 2q_{1x}a_{2}(\textbf{q}_{1})\!-
q_{x}a_{3}(\textbf{q}_{1},\textbf{q}_{x})\right \}+\]\vspace*{-7mm}
\[+\sum\limits_{p_{x}\neq 0}^{(2\pi)}\!ik_{1x}C_{1x}(p_{x})\left [4(q_{x}\!+q_{1x}\!-p_{x})
 S_{2}^{(1)}(\textbf{q}\!-\textbf{p}_{x},\textbf{q}_{1})\,-\right.\]\vspace*{-7mm}
\[-\left. 6p_{x}S_{3}^{(1)}(\textbf{q}-\textbf{p}_{x},\textbf{q}_{1},-\textbf{q}-\textbf{q}_{1})
\right ] + (x\rightarrow y, z)\,+\]\vspace*{-7mm}
\[+\frac{1}{\sqrt{N}}\sum\limits_{\textbf{q}_{2}\neq 0}^{(\pi)}\Bigg
\{\!
\frac{2}{\sqrt{N}}(q_{2}^{2}-\textbf{q}_{1}\textbf{q}_{2})S_{3}^{(1)}(\textbf{q},
\textbf{q}_{1}-\textbf{q}_{2},\textbf{q}_{2})\,+
\]\vspace*{-7mm}
\[+\frac{4}{\sqrt{N}}\textbf{q}_{2}(\textbf{q}_{1}+\textbf{q}_{2}+\textbf{q})S_{3}^{(1)}
(\textbf{q},\textbf{q}_{1},\textbf{q}_{2})\,-\]\vspace*{-7mm}
\[-\, 4\textbf{q}_{2}(\textbf{q}_{1}-\textbf{q}_{2}+\textbf{q})S_{1}^{(1)}
(\textbf{q}_{2})S_{2}^{(1)}(\textbf{q}-\textbf{q}_{2},\textbf{q}_{1})\,+\]\vspace*{-7mm}
\[+\,6q_{2}^{2}S_{1}^{(1)}(\textbf{q}_{2})S_{3}^{(1)}(\textbf{q}-
\textbf{q}_{2},\textbf{q}_{1},-\textbf{q}-\textbf{q}_{1})\,+\]\vspace*{-9mm}
\begin{equation}\label{1-35}
+\, 4(\textbf{q}_{1}+\textbf{q}_{2})^{2}S_{2}^{(1)}
(\textbf{q}_{2},\textbf{q}_{1})S_{2}^{(1)}(\textbf{q}-\textbf{q}_{2},\textbf{q}_{1}+\textbf{q}_{2})\!
\Bigg \}\!.
\end{equation}
For brevity, we do not give the equations for the higher functions $S_{j\geq 3}^{(1)}(\textbf{q})$. Equation (\ref{1-35})
should be symmetrized to satisfy (\ref{1-32b}).
For the validity of relation (\ref{1-27}), it is sufficient to
consider only $S_{j}^{(1)}$ with the first even argument. We have the rule: the first argument of  $S_{j}^{(1)}$
is quantized by the even law (\ref{1-7}). Otherwise, we set $S_{j}^{(1)}= 0$.

Analogously, (\ref{1-26}) yields the chain of equations for $\tilde{E}^{b}_0$ and the functions $a_{j}$:
\begin{equation}\label{1-4b}
 \tilde{E}^{b}_0 = \frac{N-1}{2^{d+1}N}n\nu_{3}(0) -
 \sum\limits_{\textbf{k}\not= 0}^{(\pi)}\frac{n\nu_{3}(k)}{2^{d+1}N} -
 \sum\limits_{\textbf{k}\not= 0}^{(\pi)}\frac{\hbar^{2}k^2}{2mN}a_{2}(\textbf{k}),
 \end{equation}\vspace*{-9mm}
\[\frac{n\nu_{3}(k)m}{2^{d}\hbar^2} + a_{2}(\textbf{k})k^2   -
 a_{2}^{2}(\textbf{k})k^2 = A_{2}(\textbf{k})\,+\]\vspace*{-7mm}
\begin{equation}\label{1-5b}
+\sum\limits_{\textbf{q}\not= 0}^{(\pi)}\frac{a_{3}(\textbf{k},\textbf{q})}{N}
 (q^2+\textbf{k}\textbf{q}) +
  \sum\limits_{\textbf{q}\not= 0}^{(\pi)} q^2
\frac{a_{4}(\textbf{q},-\textbf{q},\textbf{k})}{2N},
\end{equation}\vspace*{-7mm}
\begin{equation}\label{1-a3b}
 a_{3}(\textbf{k}, \textbf{q}) \approx A_{3}(\textbf{k},\textbf{q})
 -\frac{2R(\textbf{k},\textbf{q})}{\epsilon_{0}(k)+\epsilon_{0}(q)+
 \epsilon_{0}(\textbf{k}+\textbf{q})}.
 \end{equation}
 These equations can be found from (\ref{1-4})--(\ref{1-a3r})
by the changes $\nu_{3}(k)\rightarrow \nu_{3}(k)/2^d$,  $\sum\limits^{(2\pi)}\rightarrow \sum\limits^{(\pi)}$ and the addition of
  ``surface'' corrections  $A_{j}$.
The latter are related to the appearance of the terms
\[ -\frac{\hbar^2}{2m}\sum\limits_{\textbf{k}\neq 0}^{(\pi)}A_{2}(\textbf{k})\rho_{\textbf{k}}\rho_{-\textbf{k}}, \]\vspace*{-7mm}
\[ -\frac{\hbar^2}{2m}\sum\limits_{\textbf{k}_{1},\textbf{k}_{2}\neq 0}^{(\pi)
 \textbf{k}_{1}+\textbf{k}_{2}\neq 0}A_{3}(\textbf{k}_{1},\textbf{k}_{2})
 \rho_{\textbf{k}_{1}}\rho_{\textbf{k}_{2}}\rho_{-\textbf{k}_{1}-\textbf{k}_{2}}, ... \]
in Eq. (\ref{1-27}), which have the structure of terms of Eq. (\ref{1-26}).
 In particular,\vspace*{-2mm}
\[A_{2}(\textbf{k}) = \sum\limits_{q_{x}\neq 0}^{(2\pi)}\frac{ik_{1x}C_{1x}(-q_{x})}{\sqrt{N}}\left
[4(q_{x}+k_{x})S_{2}^{(1)}(\textbf{q}_{x},\textbf{k})\,+\right.\]\vspace*{-7mm}
\[+\left. 6q_{x}S_{3}^{(1)}(\textbf{q}_{x},-\textbf{q}_{x},\textbf{k})
\right ] +(x \rightarrow y, z)\,+\]\vspace*{-8mm}
\[+ \frac{1}{N}\sum\limits_{\textbf{q}\neq 0}^{(2\pi)}\left [4(q^{2}+\textbf{q}\textbf{k})S_{1}^{(1)}(-\textbf{q})
S_{2}^{(1)}(\textbf{q},\textbf{k})\,+ \right.\]\vspace*{-7mm}
\[+ \,4(\textbf{k}+\textbf{q})^{2}S_{2}^{(1)}(\textbf{q},\textbf{k})S_{2}^{(1)}(-\textbf{q},-\textbf{k})\,+\]\vspace*{-8mm}
\begin{equation} \label{1-36}
 + \left. 6q^{2}S_{1}^{(1)}(-\textbf{q}) S_{3}^{(1)}(\textbf{q},\textbf{k},-\textbf{k})\right ]\!.
\end{equation}
 Though the majority of terms in
(\ref{1-26}) and (\ref{1-27}) has the structure of the collections $\rho_{\textbf{k}}$,
 which is inherent only in the given equation ((\ref{1-26}) or (\ref{1-27})),
 a small part of terms in (\ref{1-27}) has the structure of terms in (\ref{1-26}).
 The criterion for the construction of the chains of equations
 for  $S_{j}^{(1)}$ and $a_{j}$ is the functional difference  of the collections $\rho_{\textbf{k}}$ (\ref{1-op}).
 Basically, the structures of the terms in (\ref{1-26}) and (\ref{1-27}) are different. Therefore, the partition  of Eq. (\ref{1-15}) into (\ref{1-26}) and (\ref{1-27})
 is justified and simplifies the determination of $S_{w}^{(1)}$. The collections $\rho_{\textbf{k}}$, which are present in both
equations, should be analyzed with the use of the initial equation
 (\ref{1-15}).
We can verify that $\Psi_{0}$ (\ref{1-23}) satisfies Eq. (\ref{1-15}),
if $a_{j}$ and $A_{j}$ satisfy Eqs. (\ref{1-4b})--(\ref{1-36}).

 As $U_{s}\rightarrow \infty$ and $\delta \rightarrow 0,$ relations
(\ref{1-23}), (\ref{1-32}), (\ref{1-33})--(\ref{1-36}) are the
\textit{required solution} for the WF of the ground state of $N$
interacting Bose particles filling a rectangular vessel.

The analysis of Eqs. (\ref{1-34}) and (\ref{1-35}) indicates that
the functions $S_{1}^{(1)}(\textbf{q})$ and
$S_{2}^{(1)}(\textbf{q},\textbf{q}_{1})$ can be considered
``one-dimensional'':\vspace*{-2mm}
\begin{equation} \label{1-s1}
 S_{1}^{(1)}(\textbf{q}) \!\approx\! \delta_{\textbf{q},\textbf{q}_{x}}S_{1}^{(1)}(\textbf{q}_{x})+
   \delta_{\textbf{q},\textbf{q}_{y}}S_{1}^{(1)}(\textbf{q}_{y}) +\delta_{\textbf{q},\textbf{q}_{z}}S_{1}^{(1)}(\textbf{q}_{z}),
\end{equation}\vspace*{-9mm}
\[S_{2}^{(1)}(\textbf{q},\textbf{q}_{1}) \approx  \delta_{\textbf{q},\textbf{q}_{x}}S_{2}^{(1)}
 (\textbf{q}_{x},\textbf{q}_{1})\, +\]\vspace*{-7mm}
 \begin{equation}\label{1-s2}
+\,\delta_{\textbf{q},\textbf{q}_{y}}S_{2}^{(1)}(\textbf{q}_{y},\textbf{q}_{1})
+
\delta_{\textbf{q},\textbf{q}_{z}}S_{2}^{(1)}(\textbf{q}_{z},\textbf{q}_{1})
\end{equation}
($\textbf{q}\equiv (q_{x},q_{y},q_{z})=\textbf{q}_{x}+\textbf{q}_{y}+\textbf{q}_{z}$), since the nonone-dimensional
$S_{j}^{(1)}$ with $\textbf{q}=(q_{x},q_{y},0)$ (with possible permutations of
the $x$-, $y$-, and $z$-components) and $\textbf{q}=(q_{x},q_{y},q_{z})$ are less than one-dimensional ones by,
respectively, $N^{1/3}$  and  $N^{2/3}$ times. Due to the smallness of nonone-dimensional terms, their inclusion in
the equations does not influence the values of one-dimensional ones and the value of $E_{0}$.
For the one-dimensional parts of the functions $S_{j}^{(1)},$ all terms in Eqs. (\ref{1-34}) and (\ref{1-35}) are of the same order,
as it is easy to verify. Therefore, all they should be taken into account. But, in this case, (\ref{1-33})--(\ref{1-35}) is a complicated system
of nonlinear integral equations difficult to be solved. Below, we will use the simplest zero approximations for $S_{1}^{(1)}$ and $S_{2}^{(1)}$:
\begin{equation}\label{1-38}
 S_{1}^{(1)}(\textbf{q}_{x}) \approx -\frac{i\sqrt{N}k_{1x}C_{1x}(q_{x})2a_{2}
 (-\textbf{q}_{x})}{q_{x}-2q_{x}a_{2}(\textbf{q}_{x})},
\end{equation}\vspace*{-7mm}
\[ S_{2}^{(1)}(\textbf{q}_{x},\textbf{q}_{1}) \approx\]\vspace*{-7mm}
\begin{equation}\label{1-38b}
 \approx\frac{i\sqrt{N}k_{1x}C_{1x}(q_{x})}{1-2a_{2}(\textbf{q}_{x})} \frac{2q_{1x}a_{2}(\textbf{q}_{1})-q_{x}a_{3}(\textbf{q}_{1},\textbf{q}_{x})}{\epsilon_{0}(\textbf{q}_{1})+
 \epsilon_{0}(\textbf{q}_{x}+\textbf{q}_{1})}.
\end{equation}

It is easy to verify that the surface corrections $A_{1}$ and $A_{2}$ have no effect on the solutions for $E_{0}$ and $a_{2}(k)$.
We expect that $A_{j\geq 3}$ do not affect the solutions for  $a_{j\geq 3}$.
However, the
 boundaries  change the ``bulk part'' of the equations for $E_{0}$ and $a_{j}$:
the network of vectors $\textbf{k}$ becomes by a factor of $2^d$ denser, and the Fourier-transform of the potential is multiplied by $1/2^d$.
For this reason, the boundaries strongly influence the values of
$E_{0}$ and $a_{j}$.

Besides WF (\ref{1-23}), WFs of the form (\ref{1-23}) with
$\Psi^{f}_{\textbf{0}} \rightarrow $
$\rightarrow\Psi^{f}_{\textbf{l}}$  (and $\textbf{k}_{1}\rightarrow
\textbf{k}_{\textbf{l}}$) are also the solutions. But such WFs are
equal to zero at many points inside the vessel.

\section{Ground State of He II: Solutions}

The sines
standing before the exponential function in $\Psi_{0}$ (\ref{1-23}) set an
inhomogeneous arc-like distribution of atoms in the vessel.
For this reason, the anxieties arose \cite{bijl,gir1965,huang} that the use of bare
sines will not allow one to construct the series of perturbation theory so that
a reasonable behavior of the density (the constant density of He II in the whole volume
except for a narrow strip near the wall) be obtained. Apparently, namely for this reason,
the problem was solved with periodic BCs, rather than with zero BCs;
and the above-obtained solution was lost.
Therefore, it is important to verify solution (\ref{1-23}) in this aspect.
 Let us find the behavior of
the density.  Since the function $e^{\tilde{S}_{b}}$ sets a
homogeneous distribution of atoms, we need to consider the
behavior of the ansatz $\Psi^{f}_{\textbf{0}}e^{S_{w}^{(1)}}$ in
 (\ref{1-23}). The function
$S_{w}^{(1)}(\textbf{r}_{1},... ,\textbf{r}_{N})$ (\ref{1-32})
contains the one-particle dependence in each sum, and the values are
of the same order. But taking the  sums with $S_{j\geq
2}^{(1)}$ into account is a  complicated task, and we neglect it. Then
\begin{equation}\label{1-37a}
\Psi_{0}(x_{1})\sim f(\textbf{r}_{1}) = \sin{(k_{1x}x_{1}+\delta_{1x})}e^{s_{1}(\textbf{r}_{1})},
\end{equation}\vspace*{-7mm}

\begin{figure}%
\vskip1mm
\includegraphics[width=5cm]{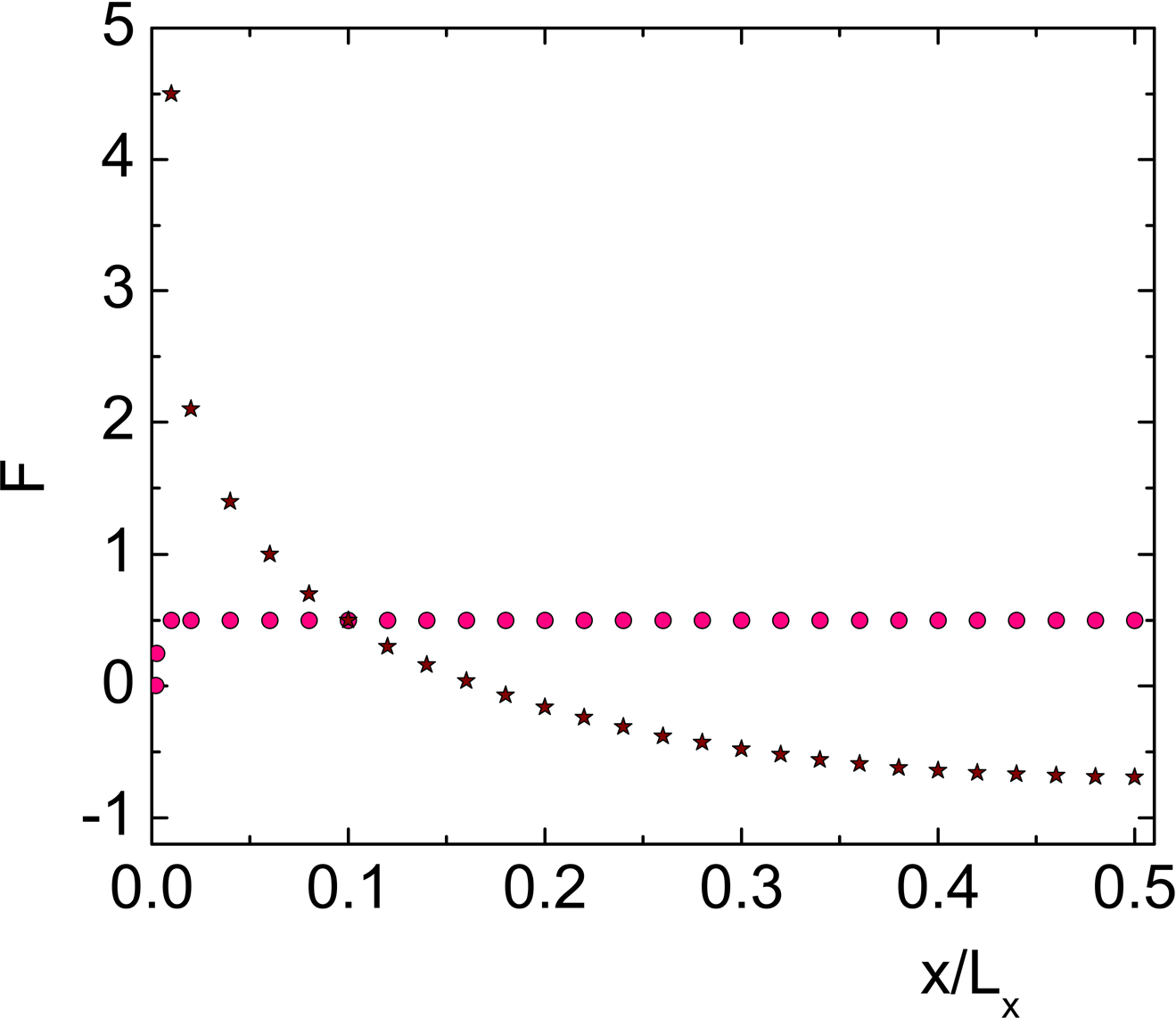}
\vskip-3mm\caption{ Values of the functions $f(x)$ ($\circ$) and
$s_{1}(x)$ ($\star$) by formulas (\ref{1-37a}), (\ref{1-37}) for
potentials (\ref{1-10}) and (\ref{1-39}) at $a=2\,\mbox{\AA}$,
$U_{b} =1600\,$K, $U_{bd} =0\,$K,  $U_{s} \rightarrow \infty,$ and
the vessel size $L_{x}=
10^{4}\bar{R}$. Values of
$f(x)$ and $s_{1}(x)$ at  $x/L_{x} \in [0.5, 1]$ are obtained by the
reflection of the curves in an upright mirror positioned at
$x=0.5L_{x}$ }\label{fig1}
\end{figure}

\begin{figure}%
\vskip3mm
\includegraphics[width=5cm]{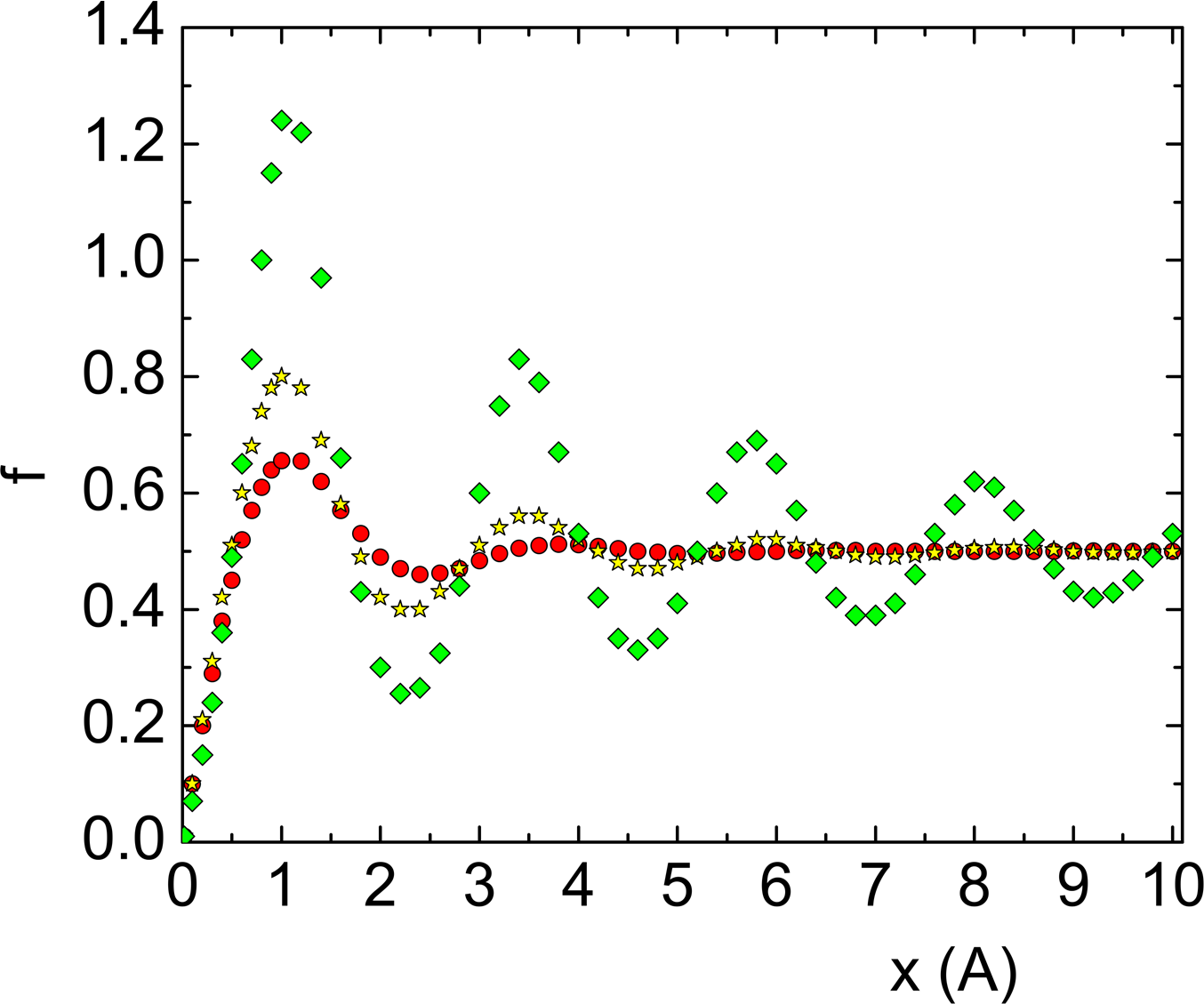}
\vskip-3mm\caption{ Values of the function $f(x)$ (\ref{1-37a}),
(\ref{1-37}) near the wall ($x=0$) at $L_{x}=10^{4}\bar{R}$, $U_{s}
\rightarrow \infty,$ and potential (\ref{1-39}) with
$a=2\,\mbox{\AA}$, $b=4\,\mbox{\AA}$, $U_{bd} =-9\,$K, and three
barriers: $U_{b} =800\,$K ($\circ$), $U_{b} =2000\,$K ($\star$), and
$U_{b} =2900\,$K ($\diamond$). At $x\gsim 6\,\mbox{\AA},$ curves
$\circ\circ\circ$ and $\star\star\star$ almost merge. Values of $x$
are given in {\AA} }\label{fig2}
\end{figure}

\noindent\[s_{1}(\textbf{r}_{1})=\sum\limits_{q_{x}\neq 0}^{(2\pi)}
\frac{S_{1}^{(1)}
(\textbf{q}_{x})}{\sqrt{N}}e^{iq_{x}x_{1}}\,+\]\vspace*{-6mm}
\[+\sum\limits_{q_{x},q_{y}\neq 0}^{(2\pi)}\frac{S_{1}^{(1)}(\textbf{q}_{x}+\textbf{q}_{y})}
{\sqrt{N}}e^{iq_{x}x_{1}+iq_{y}y_{1}}\,+\]\vspace*{-6mm}
\[+\sum\limits_{q_{x},q_{z}\neq 0}^{(2\pi)}\frac{S_{1}^{(1)}(\textbf{q}_{x}+\textbf{q}_{z})}
{\sqrt{N}}e^{iq_{x}x_{1}+iq_{z}z_{1}}\,+\]\vspace*{-6mm}
\begin{equation} \label{1-37b}
+\!\!\!\sum\limits_{q_{x},q_{y},q_{z}\neq
0}^{(2\pi)}\!\!\frac{S_{1}^{(1)}
(\textbf{q}_{x}+\textbf{q}_{y}+\textbf{q}_{z})}{\sqrt{N}}e^{iq_{x}x_{1}+iq_{y}y_{1}+iq_{z}z_{1}}.
\end{equation}

\noindent As was noted in Sec. 2, we have
$S_{1}^{(1)}(\textbf{q}_{x}+\textbf{q}_{y}+\textbf{q}_{z}) \sim$ $
\sim S_{1}^{(1)}(\textbf{q}_{x}+\textbf{q}_{y})/N^{1/3}\sim
S_{1}^{(1)}(\textbf{q}_{x})/N^{2/3}.$  Therefore, three last sums in
(\ref{1-37b}) are of the order of the first one, but they depend
additionally on $y_{1}$ and $z_{1}$. Let us consider only the first
sum:\vspace*{-1mm}
\begin{equation}
 s_{1}(\textbf{r}_{1})=
 s_{1}(x_{1})=\sum\limits_{q_{x}=2\pi j_{x}/L_{x}}^{(2\pi)}\frac{S_{1}^{(1)}(\textbf{q}_{x})}{\sqrt{N}}e^{iq_{x}x_{1}},
      \label{1-37} \end{equation}\vspace*{-3mm}

\noindent where $j_{x}=\pm 1, \pm 2, \pm 3,  ...\,$. For
$S_{1}^{(1)}(\textbf{q}_{x}),$ we use the zero approximation
(\ref{1-38}). In view of the property
$S_{1}^{(1)}(-\textbf{q}_{x})=S_{1}^{(1)}(\textbf{q}_{x})$ and since
$S_{1}^{(1)}(\textbf{q}_{x})$ is real, the imaginary part of
(\ref{1-37}) is cancelled, and we can change
$e^{iq_{x}x_{1}}\rightarrow \cos{(q_{x}x_{1})}$. In this paper, we
study only the qualitative behavior of the solutions. Therefore, we
choose the interatomic potential of two He$^4$ atoms in the simple
form\vspace*{-1mm}
 \begin{equation}
 U_{3}(\textbf{r}) \approx
\left \{\!\! \begin{array}{ll}
    U_{b}>0,  & \   r\leq a ,  \\[1mm]
    U_{bd}<0, & \   a\leq r \leq b , \\[1mm]
    0,  & \ r>b.
\label{1-39} \end{array} \right.
\end{equation}\vspace*{-3mm}

\noindent For the modern potentials \cite{aziz1997,szalewicz2012},
$a\approx 2.6\,\mbox{\AA}$,  $U_{bd}\approx$\linebreak
$\approx-11\,$K, but the barrier and the well are more flat than
steps (\ref{1-39}). Therefore, we will use the lower values of $a$
and $|U_{bd}|$ in (\ref{1-39}): $a= 2\,\mbox{\AA}$,  $U_{bd}=-9\,$K,
and $b= 4\,\mbox{\AA}$.

The Fourier-transform (\ref{1-6}) of potential (\ref{1-39})
is\vspace*{-2mm}
\[ \nu_{3}(k)  = \frac{4\pi }{k^3}\left \{
       [U_{b}-U_{bd}][\sin{(ak)}-ak\cos{(ak)]}\right.+\]\vspace*{-8mm}
\begin{equation}\label{1-40}
 + \left. U_{bd}[\sin{(bk)}-bk\cos{(bk)}]\right \}\!.
\end{equation}
In the two-particle approximation, we have from
(\ref{1-5b}):\vspace*{-1mm}
\begin{equation}\label{1-41}
a_{2}(k) = \frac{1}{2}- \sqrt{\frac{1}{4}+\frac{n \nu_{3}(k)m}{2^{d}\hbar^{2}k^{2}}}.
\end{equation}
 We chose the
sign ``minus'' before the root in Eq.\,(\ref{1-41}). At the sign
``plus,'' we have nonphysical result \mbox{$E_{0}=-\infty$.}

The results for the functions  $f(x_{1})$ and $s_{1}(x_{1})$
(\ref{1-37a}), (\ref{1-37}) at $L_{x}=10^{4}\bar{R}$ and several
potentials are given in Figs. 1 and 2; here, $\bar{R}\approx
3.58\,\mbox{\AA}$ is the mean interatomic distance for He II. The
values of the functions at $L_{x}=10^{5}\bar{R}$ and
$L_{x}=10^{3}\bar{R}$ differ from the drawn ones very slightly (by
$\lsim 0.1\%$). So, the curves in the figures are true for all
$L_{x}\gsim 10^{3}\bar{R}$. As seen from Fig.~1,  $f(x_{1})$ is
equal to 0.5 everywhere, except for a narrow region near the wall.
While approaching the wall, $f(x_{1})$ oscillates with constant
period and increasing amplitude and becomes zero on the wall. An
increase of the interatomic barrier $U_{b}$ causes a growth of the
amplitude of oscillations $f(x_{1})$. The well depth $U_{bd}$ does
not affect the results, if  $|U_{bd}|\ll U_{b}$. As the distance to
the wall increases, $f(x_{1})$ approaches the constant $0.5$ with a
high accuracy increasing with $L_{x}$. For example, at
$L_{x}=10^{4}\bar{R},$ the deviation from 0.5 occurs only in the
sixth decimal point. As $x_{1}\rightarrow L_{x},$ the function
$f(x_{1})$ behaves itself as at $x_{1}\rightarrow 0$.

It is of interest that $f(x_{1})$ oscillates near the wall. This is
due to the short-range order, and the period is proportional to
$a/\bar{R}$. Large oscillations at $U_{b} =$ $=2900\,$K are, most
likely, nonphysical and are related to the neglect of corrections,
which are large at such a potential. This  reflects the behavior of
the density $\rho$ of helium, since  $\rho(x_{1})\sim \int
dy_{1}dz_{1}d\textbf{r}_{2}... d\textbf{r}_{N}|\Psi_{0}|^{2}\sim$
$\sim f^{2}(x_{1})$ (the behavior of $\rho(x_{1})$ near the wall is
also affected by  $e^{\tilde{S}_{b}}$, but we did not study
it).\,\,Similar oscillations were obtained in \cite{krotscheck1988},
while modeling the properties of a film of helium adsorbed on the
\mbox{substrate.}

We obtained the solutions also at  finite $U_{s}$. At $U_{s} \gsim
1000\,$K, they  are practically invariable with increase in $U_{s}$,
except for the values on a wall, where  $f \neq 0$. As $U_{s}$
increases, $s_{1}(x)$ approaches a constant ($\simeq 10$) at $x=0$,
and $\sin{(k_{1x}x+\delta_{1x})}\rightarrow 0.$ As a result, we have
$f(x\rightarrow 0)|_{U_{s}\rightarrow \infty} \rightarrow 0$. With
regard for the higher corrections, such behavior will be
conserved:  as  $U_{s}\rightarrow \infty,$  we have $\delta
\rightarrow 0,$ such $\delta$ affects slightly $C_{1}(q)$ and all
results, because $\delta$ enters the equations through the function
$C_{1}(q)$. In addition, the terms with large $q$ are cut in sums by
$a_{j}.$ Due to such properties, the transition $U_{s}\rightarrow
\infty$, $\delta \rightarrow 0$ is correct and makes $\Psi_{0}$ to
be arbitrarily close to zero on the walls.

Thus, though the bare  function $\sin{(k_{1x}x+\delta_{1x})}$ is far
from to be a constant, the account for $S_{1}^{(1)}(q_{x})$ makes
the theoretical density of atoms  to be constant everywhere, except
for a narrow ($\sim$$\mbox{\AA}$) region near walls. Sines are the
best bare functions, because 1) their use gives the simplest
equations;
 2)~if the interatomic interaction is
``switched-off,'' $\Psi_{0}$ is reduced to the product of bare sines, which is the ground-state WF of $N$
free Bose particles in the box.
   The circumstance that the
correction $S_{1}^{(1)}(q_{x})$ ``improves'' the behavior of
$f(x_{1})$ seems natural, because the Schr\"{o}dinger equation
(\ref{1-15}) involves the interaction of atoms and must make their
distribution uniform, but  $S_{1}^{(1)}(q_{x})$ is
 generated just  by the Schr\"{o}dinger \mbox{equation.}

  \section{Excited State of He II}
 We now find the WF of helium II in a vessel for the state with one phonon. Under periodic BCs, the exact solution
for the WF takes the form \cite{yuv2}
\begin{equation}\label{4-1}
\Psi_{\textbf{k}}(\textbf{r}_1,... ,\textbf{r}_N) =
\psi_{\textbf{k}}\Psi_0,
\end{equation}\vspace*{-7mm}
\[\psi_{\textbf{k}}  \equiv  \psi^{b}_{\textbf{k}} = \rho_{-\textbf{k}} +
 \sum\limits_{\textbf{k}_{2}\neq 0, -\textbf{k}}^{(2\pi)}
  \frac{b_{2}(\textbf{k}_{2},\textbf{k})}{\sqrt{N}}
 \rho_{\textbf{k}_{2}}\rho_{-\textbf{k}_{2}-\textbf{k}}\,+ \]\vspace*{-5mm}
\begin{equation}\label{4-2}
+ \sum\limits_{\textbf{k}_{2},\textbf{k}_{3}\neq 0}^{(2\pi)\,\textbf{k}_{2}+
 \textbf{k}_{3}+\textbf{k} \not= 0}
  \frac{b_{3}(\textbf{k}_{2},\textbf{k}_{3},\textbf{k})}{N}
 \rho_{\textbf{k}_{2}}\rho_{\textbf{k}_{3}}\rho_{-\textbf{k}_{2}-\textbf{k}_{3}-\textbf{k}}
 +...\,.\vspace*{-11mm}
\end{equation}\vspace*{-6mm}

\noindent A solution in the zero approximation
\begin{equation}\label{4-3}
\psi^{b}_{\textbf{k}}  =  \rho_{-\textbf{k}}
\end{equation}
with a simplified first correction with $b_{2}$ was first proposed by
R. Feynman \cite{fey1,fc}. The arguments in favor of solution
(\ref{4-1}), (\ref{4-2}) are as follows: 1) it
satisfies  the Schr\"{o}dinger equation for $N$ interacting Bose particles and
 2) it is an eigenfunction of the momentum operator
$\hat{\textbf{P}}=-i\hbar\sum_{j}\nabla_{\textbf{r}_{j}}$ with the
eigenvalue $\hbar\textbf{k}$; 3) at the ``switching-off'' of the
interaction, it is reduced to WF (\ref{4-3}) describing $N$ free
Bose particles, from which $N-1$ particles are immovable, and the
single one  moves with the momentum $\textbf{k}$; 4) an analogous
solution is obtained in the operator
 approach \cite{bz}; 5)  the completeness
 \cite{yuv1} of collections (\ref{1-op}) implies that the solution for $\psi_{\textbf{k}}$ should have only such a form.
However, the structure of (\ref{4-1}) is  the assumption, though no solution with another structure was found.
The solution for $\psi_{\textbf{k}}$ with ``shadow'' variables \cite{swf1994a,swf1994b}
is equivalent to (\ref{4-1}), (\ref{4-2}).

The simplest way to find solution (\ref{4-1}), (\ref{4-2}) is to
assume the structure of (\ref{4-1}) and to substitute the WF of free
particles (\ref{4-3}), as the bare  $\psi_{\textbf{k}}$, in
the Schr\"{o}dinger equation for interacting particles.
The equation prompts the form of corrections to $\psi_{\textbf{k}}$.

In a similar manner, we start for the liquid in a vessel from
(\ref{4-1}) with $\Psi_0$ (\ref{1-23}). The study of different
possibilities has shown that $\psi_{\textbf{k}}$ must be sought in
the form\vspace*{-1mm}
\[\psi_{\textbf{k}_{\textbf{r}}} = \frac{8}{\sqrt{N}} \sum\limits_{j=1}^{N}
\cos{(k_{r_{x}}x_{j})} \cos{(k_{r_{y}}y_{j})}
 \cos{(k_{r_{z}}z_{j})} \,+\]\vspace*{-5mm}
\begin{equation}\label{4-6}
+\,\delta \psi_{\textbf{k}_{\textbf{r}}} =
 \rho_{-\textbf{k}_{\textbf{r}}} + 7\,\mbox{permutations}+\delta \psi_{\textbf{k}_{\textbf{r}}}
 \end{equation}
 with  the unknown $\delta \psi_{\textbf{k}_{\textbf{r}}}.$
The permutation means $\rho_{-\textbf{k}_{\textbf{r}}}$ with the different sign of one or several components of
$\textbf{k}_{\textbf{r}}$.  In this case,
 $\textbf{k}_{\textbf{r}}$ takes values
(\ref{1-11}), since we must obtain the solution
for free particles at the switching-off of the interaction.
We now represent $\psi_{\textbf{k}}$ in the form
 \begin{equation}\label{4-8}
 \psi_{\textbf{k}}  = \psi^{0}_{\textbf{k}} + 7\,\mbox{permutations},
 \end{equation}\vspace*{-7mm}
 \begin{equation}\label{4-9}
 \psi^{0}_{\textbf{k}} = \rho_{-\textbf{k}} + \delta
 \psi^{0}_{\textbf{k}}.
 \end{equation}
Solution (\ref{4-8}), (\ref{4-9}) describes a 3D
 standing wave decaying into eight counterrunning waves.

Substituting WF (\ref{4-1}) in the  Schr\"{o}dinger equation with
 Hamiltonian  (\ref{1-8}), we obtain the equa- \mbox{tions for
 $\psi_{\textbf{k}}$:}
\begin{equation}\label{4-10}
 -\frac{\hbar^{2}}{2m}\sum\limits_{j}\left \{ \triangle_{j}\psi_{\textbf{k}} +
 2(\nabla_{j}\psi_{\textbf{k}})\times \nabla_{j}S \right \} =
 E(\textbf{k})\psi_{\textbf{k}},
 \end{equation}\vspace*{-7mm}
 \begin{equation}\label{4-11}
 S=S^{(1)}_{w}+\tilde{S}_{b}+S_{0},
 \end{equation}\vspace*{-7mm}
 \begin{equation}\label{4-12}
  S_{0}=\ln{\Psi^{f}_{\textbf{0}}},
 \end{equation}
where $E(\textbf{k})=E-NE_{0}$ is the energy of a quasiparticle;
$E(\textbf{k})$ is obviously independent of the signs of components of $\textbf{k}.$ Therefore,
Eq. (\ref{4-10}) is separated into 8 equations: for
$\psi^{0}_{\textbf{k}}$ and 7 permutations with the same energy
$E(k)$.

The analysis indicates that the solution for $\psi^{0}_{\textbf{k}}$ has the following
form:
\[\psi^{0}_{\textbf{k}}  =  \psi^{b}_{\textbf{k}} + b_{0}(\textbf{k}) +
 \sum\limits_{\textbf{q} \neq 0, -\textbf{k}}^{(2\pi)}
  Q_{1}(\textbf{q}, \textbf{k})\rho_{-\textbf{q}-\textbf{k}} \,+\]\vspace*{-7mm}
\[+ \sum\limits_{\textbf{q}, \textbf{q}_{1}\neq 0}^{\textbf{q}+\textbf{q}_{1}+
 \textbf{k} \not= 0}
  \frac{Q_{2}(\textbf{q},\textbf{q}_{1},\textbf{k})}{\sqrt{N}}
 \rho_{\textbf{q}_{1}}\rho_{-\textbf{q}-\textbf{q}_{1}-\textbf{k}}\,+\]\vspace*{-5mm}
\[ + \sum\limits_{\textbf{q}, \textbf{q}_{1}, \textbf{q}_{2}\neq 0}^{\textbf{q}+\textbf{q}_{1}+\textbf{q}_{2}
 +\textbf{k} \not= 0}
  \frac{Q_{3}(\textbf{q},\textbf{q}_{1},\textbf{q}_{2},\textbf{k})}{N}
 \rho_{\textbf{q}_{1}}\rho_{\textbf{q}_{2}}\rho_{-\textbf{q}-\textbf{q}_{1}-\textbf{q}_{2}-\textbf{k}}\,
 +\]\vspace*{-7mm}
\begin{equation}\label{4-13}
+... ,
\end{equation}
where  $\psi^{b}_{\textbf{k}}$ is of the form of solution (63) for
periodic BCs (but with $\textbf{k}_{j}$ running series
(\ref{1-11})),   $b_{0}$ and $Q_{l}$ are the corrections from
boundaries. Here,  $\textbf{q}$ is quantized according to
(\ref{1-7}) (like $2\pi j/L$), and   $\textbf{q}_{j}$ and
$\textbf{k}$ are quantized by (\ref{1-11})
 (like $\pi j/L$).

With regard for the properties of
 $\rho_{\textbf{k}}$, we can directly verify that $\Psi_{\textbf{k}}$  (\ref{4-1}), (\ref{4-8}), (\ref{4-13})
 is the exact solution of the Schr\"{o}dinger equation, if  the functions
 $E(k)$,  $b_{j},$ and $Q_{l}$ satisfy  the equations
\[b_{0}(\textbf{k})\epsilon(k)  =  -2k^{2}S_{1}^{(1)}(-\textbf{k})\delta_{\textbf{k},\textbf{k}^{e}}\,-\]\vspace*{-7mm}
\[-\, 2k_{1x}k_{x}iC_{1x}(-k_{x})\sqrt{N}\delta_{\textbf{k},\textbf{k}_{x}^{e}}\,-\]\vspace*{-9mm}
\[ -\, \frac{2}{\sqrt{N}}\sum\limits_{\textbf{q}\neq 0}^{(\pi)}q^{2}Q_{2}
    (-\textbf{k},\textbf{q},\textbf{k})\delta_{\textbf{k},\textbf{k}^{e}} \,-\]\vspace*{-7mm}
\[ - \sum\limits_{\textbf{q}\neq 0, -\textbf{k}}^{(2\pi)}2(\textbf{k}+\textbf{q})^{2}Q_{1}(\textbf{q},\textbf{k})
  S_{1}^{(1)}(-\textbf{k}-\textbf{q})\delta_{\textbf{k},\textbf{k}^{e}}\,+\]\vspace*{-7mm}
\[ + \sum\limits_{q_{x}\neq 0}^{(2\pi)}2\sqrt{N}k_{1x}q_{x}iC_{1x}(q_{x})Q_{1}
 (-\textbf{k}-\textbf{q}_{x},\textbf{k})\delta_{\textbf{k},\textbf{k}^{e}}\,  +\]\vspace*{-7mm}
 \begin{equation}\label{4-14c}
+\, (x\rightarrow y, z),
\end{equation}\vspace*{-7mm}
\[ \epsilon(k)  =  \epsilon_{0}(k)-\frac{1}{N}\sum\limits_{\textbf{k}_{2}\neq 0}^{(\pi)}
 b_{2}(\textbf{k}_{2}, \textbf{k})2\textbf{k}_{2}(\textbf{k}_{2}+\textbf{k})\,-\]\vspace*{-7mm}
\[-\frac{1}{N}\sum\limits_{\textbf{k}_{2}\neq 0}^{(\pi)}
 6k^{2}_{2} b_{3}(\textbf{k}_{2}, -\textbf{k}_{2}, \textbf{k})\,-\]\vspace*{-7mm}
\[- \sum\limits_{q_{x}\neq 0}^{(2\pi)} 2k_{1x}(k_{x}+q_{x})iC_{1x}(-q_{x})Q_{1}(\textbf{q}_{x},\textbf{k})\, -\]\vspace*{-7mm}
\[- \frac{2}{\sqrt{N}}\sum\limits_{\textbf{q}\neq 0}^{(2\pi)} Q_{1}
(\textbf{q},\textbf{k})(\textbf{k}+\textbf{q})\,\times\]\vspace*{-7mm}
\[ \times \left [ \textbf{q}S_{1}^{(1)}(-\textbf{q})+2(\textbf{k}+\textbf{q})S^{(1)}_{2}(-\textbf{q}, -\textbf{k})
  \right ]+\]\vspace*{-7mm}
\[+\sum\limits_{q_{x}\neq 0}^{(2\pi)} 4k_{1x}q_{x}iC_{1x}(q_{x})Q_{2}(-\textbf{q}_{x},\textbf{q}_{x},\textbf{k})\,-\]\vspace*{-7mm}
\begin{equation}\label{4-14}
- \frac{4}{\sqrt{N}}\sum\limits_{\textbf{q}\neq 0}^{(2\pi)} q^{2}Q_{2}(-\textbf{q},\textbf{q},\textbf{k})
S_{1}^{(1)}(\textbf{q})+ (x\rightarrow y, z),
\end{equation}\vspace*{-7mm}
\[Q_{1}(\textbf{q},\textbf{k})\left
[\epsilon(k)-\epsilon_{0}(\textbf{k}+\textbf{q})
 \right ]=\]\vspace*{-7mm}
\[=\,  2ik_{1x}C_{1x}(q_{x})\delta_{\textbf{q},\textbf{q}_{x}}\left [
 -k_{x}+2q_{x}b_{2}(\textbf{q}, \textbf{k})
  \right ]+\]\vspace*{-7mm}
\[+ \frac{2S_{1}^{(1)}(\textbf{q})}{\sqrt{N}}[\textbf{q}\textbf{k}-2q^{2}b_{2}(\textbf{q}, \textbf{k})]
  -\frac{4k^2}{\sqrt{N}}S_{2}^{({1})}(\textbf{q},-\textbf{q}-\textbf{k}) \,-\]\vspace*{-7mm}
\[- \frac{1}{N}\sum\limits_{\textbf{q}_{1}\neq 0}^{(\pi)} \left \{ 2\textbf{q}_{1}(\textbf{q}_{1}+\textbf{q}+
  \textbf{k})\right.\times\]\vspace*{-7mm}
\[\times\, [Q_{2}(\textbf{q},\textbf{q}_{1},\textbf{k})+Q_{1}(\textbf{q}
+\textbf{q}_{1},\textbf{k})\sqrt{N}S_{1}^{(1)}(-\textbf{q}_{1})
]\,+\]\vspace*{-7mm}
\[+\,
6q_{1}^{2}Q_{3}(\textbf{q},\textbf{q}_{1},-\textbf{q}_{1},\textbf{k})\,+\]\vspace*{-7mm}
\[+\, 4q_{1}^{2}Q_{2}(\textbf{q}-\textbf{q}_{1},\textbf{q}_{1},\textbf{k})\sqrt{N}S_{1}^{(1)}(\textbf{q}_{1})+
  Q_{1}(\textbf{q}+\textbf{q}_{1},\textbf{k})\,\times\]\vspace*{-7mm}
\[ \times  \left. 4(\textbf{q}_{1}+\textbf{q}+ \textbf{k})^{2}\sqrt{N}S_{2}^{(1)}
   (-\textbf{q}_{1},-\textbf{q}-\textbf{k}) \!\right \}+\]\vspace*{-7mm}
\[+\!\! \sum\limits_{p_{x}\neq 0}^{(2\pi)} 2k_{1x}iC_{1x}(p_{x})
  \left [(-k_{x}-q_{x}+p_{x})Q_{1}(\textbf{q}-\textbf{p}_{x},\textbf{k})\right.+\]\vspace*{-7mm}
\begin{equation}\label{4-15}
+ \left. 2p_{x}
Q_{2}(\textbf{q}-\textbf{p}_{x},\textbf{p}_{x},\textbf{k}) \right ]+(x\rightarrow y, z),
\end{equation}\vspace*{-7mm}
\[b_{2}(\textbf{k}_{2}, \textbf{k})\left
[\epsilon(k)-\epsilon_{0}(\textbf{k}_{2})-\epsilon_{0}(\textbf{k}+\textbf{k}_{2})
 \right ]=\]\vspace*{-7mm}
\[= \textbf{k}\textbf{k}_{2}a_{2}(\textbf{k}_{2})-\textbf{k}(\textbf{k}+\textbf{k}_{2})
 a_{2}(\textbf{k}+\textbf{k}_{2})-k^{2}a_{3}(\textbf{k}, \textbf{k}_{2})\,-\]\vspace*{-7mm}
\[ - \frac{2}{\sqrt{N}}\sum\limits_{\textbf{k}_{3}\neq 0}^{(\pi)} \bigg \{\! \frac{3\textbf{k}_{3}
 (\textbf{k}_{2}+\textbf{k}_{3}+
  \textbf{k})}{\sqrt{N}}b_{3}(\textbf{k}_{2},\textbf{k}_{3},\textbf{k})\,+\]\vspace*{-7mm}
\[+\,Q_{1}(\textbf{k}_{3},\textbf{k})(\textbf{k}_{3}+\textbf{k})\left [2(\textbf{k}_{3}
-\textbf{k}_{2})S_{2}^{(1)}(-\textbf{k}_{3},\textbf{k}_{2}) \right.
+\]\vspace*{-7mm}
\[+\left. 3(\textbf{k}_{3}+\textbf{k})S_{3}^{(1)}(-\textbf{k}_{3},\textbf{k}_{2},
-\textbf{k}_{2}-\textbf{k})\right ] +\]\vspace*{-7mm}
\[+\, 2Q_{2}(-\textbf{k}_{3},\textbf{k}_{2}+\textbf{k}_{3},\textbf{k})(\textbf{k}_{2}+\textbf{k}_{3}) \,\times\]\vspace*{-7mm}
\[\times \left [\textbf{k}_{3}S_{1}^{(1)}(\textbf{k}_{3})+2(\textbf{k}_{2}+\textbf{k}_{3})S_{2}^{(1)}
(\textbf{k}_{3},\textbf{k}_{2}) \right ]+\]\vspace*{-7mm}
\[+ \, 3k_{3}^{2}S_{1}^{(1)}(\textbf{k}_{3})Q_{3}(-\textbf{k}_{3},\textbf{k}_{2},
-\textbf{k}_{2}-\textbf{k},\textbf{k})\! \bigg \}
\,+\]\vspace*{-7mm}
\[+ \sum\limits_{p_{x}\neq 0}^{(2\pi)} 2k_{1x}iC_{1x}(p_{x})
  \left [3p_{x}Q_{3}(-\textbf{p}_{x},\textbf{k}_{2},-\textbf{k}_{2}-\textbf{k},\textbf{k})\right. +\]\vspace*{-7mm}
\[+ \left. 2(k_{2x}+p_{x})
  Q_{2}(-\textbf{p}_{x},\textbf{k}_{2}+\textbf{p}_{x},\textbf{k}) \right ]+\]\vspace*{-7mm}
\begin{equation}\label{4-16}
+\, \mbox{higher corrections} +(x\rightarrow y, z),
\end{equation}
 where $\epsilon(k)=2mE(k)/\hbar^2$, $\textbf{k}^{e}$ is the wave vector with  even components (i.e., they are multiple to $2\pi/L$),
 $\textbf{q}_{x}=q_{x}\textbf{i}_{x},
 \textbf{q}_{y}=q_{y}\textbf{i}_{y}, \textbf{q}_{z}=q_{z}\textbf{i}_{z},$ and analogously for
$\textbf{p}_{x}$, $\textbf{p}_{y}$, and $\textbf{p}_{z}.$ By
  $(x\rightarrow y, z),$ we denote
  the same terms as that with the separated $x$-component, but with the  changes $x\rightarrow y$ and $x\rightarrow z$.

Equations (\ref{4-14c})--(\ref{4-16}) are written in the approximation
of ``two sums in the wave vector'', at which the series contain the
functions $a_{2}$, $a_{3}$, $b_{2}$, $b_{3}$, $S^{(1)}_{j\leq 3}$, and $Q_{l\leq 3}$ and
do not include the corrections $a_{j\geq 4}$, $b_{j\geq 4}$, $S^{(1)}_{j\geq 4},$
 and $Q_{l\geq 4}$ (they change the equation for $b_{2}$, by adding
the correcting sums with $b_{4}$, $S^{(1)}_{4},$
 and $Q_{4}$).

If the interaction is switched-off,  $\Psi_{\textbf{k}}$
reveals a complicated structure, which should be reduced
to the solution for free particles; but we omit this point.

Let us consider the properties of Eqs. (\ref{4-13})--(\ref{4-16}).
Solution (\ref{4-2}) was obtained under periodic BCs and contains
only the functions $b_{j\geq 2}$. Solution (\ref{4-13}) corresponds
to the zero BCs and contains the additional functions  $b_{0}$ and
$Q_{l}$. In view of the ``one-dimensional''  form (\ref{1-s1}),
(\ref{1-s2}) of the functions $S_{j}^{(1)}$, relations
(\ref{4-14c})--(\ref{4-16}) imply that the functions $Q_{l}$ can
be also considered ``one-dimensional.'' In particular,
\[
 Q_{1}(\textbf{q},\textbf{k}) \approx
 Q_{1}(\textbf{q}_{x},\textbf{k})\delta_{\textbf{q},\textbf{q}_{x}}\,
 +
\]\vspace*{-7mm}
\begin{equation}\label{4-17}
   +\,Q_{1}(\textbf{q}_{y},\textbf{k})\delta_{\textbf{q},\textbf{q}_{y}}
 + Q_{1}(\textbf{q}_{z},\textbf{k})\delta_{\textbf{q},\textbf{q}_{z}}.
 \end{equation}
The ``two-dimensional'' $Q_{1}$  with $\textbf{q}$ of the form
$(q_{x},q_{y},0)$  and the ``three-dimensional'' ones with
$\textbf{q}=$ $=(q_{x},q_{y},q_{z})$ are less than the
one-dimensional ones by $\sim$$ N^{1/3}$ and $\sim$$N^{2/3}$ times,
respectively. Due to the smallness of ``non-1D'' $Q_{1},$ their
inclusion in the equations does not affect the values of
one-dimensional ones. The estimates indicate that  $Q_{l}$ affects
$\epsilon(k)$ negligibly. In Eq. (\ref{4-16}), the surface
corrections for $b_{2}$ are also less than the bulk ones by $\sim$$
N^{1/3}$ times. The smallness of surface corrections $A_{l}$ and
$Q_{l}$ is related to the thinness of a near-surface fluid layer as
compared with the sizes of a system.

Therefore, we set $Q_{l}=0$ in Eqs. (\ref{4-13})--(\ref{4-16}). Then we obtain
the Vakarchuk--Yukhnovskii equations \cite{yuv2} for $\epsilon(k)$ and $b_{j}$. However,  $\textbf{k}_{j}$
in these equations should be quantized by the  $\pi/L$-law (\ref{1-11}) instead of the usually used
 $2\pi/L$-law (\ref{1-7}), which is valid for the periodic BCs.
In the zero approximation for $\Psi_{0}$ and $\Psi_{\textbf{k}},$ we set $a_{j\geq 3}=0$ and $b_{j}=0$ in the equations.
Then we obtain
the formula for the energy of quasiparticles:
\begin{equation}\label{4-18}
 E(k) \approx \sqrt{\left (\frac{\hbar^{2} k^2}{2m}\right )^{2} +
  \frac{n\nu_{3}(k)}{2^d}\frac{\hbar^2 k^2}{m}},
\end{equation}
which is close to the Bogolyubov one, but it contains the additional factor $1/2^d$ due to
 the boundaries
(\textit{in the general case, $d$ is the number of nonperiodic
coordinates}). This formula is true at a weak interaction.
 Comparing formula (\ref{4-18}) at potential (\ref{1-39}) with the dispersion curve of He II as $k\rightarrow 0$,
we obtain $U_{b}\simeq 283\,$K, whereas we have $U_{b}\simeq 35\,$K without
the factor $1/2^d$. The first estimate is more plausible.

The Bogolyubov--Zubarev model \cite{bz} gives the same results, if
we expand  the potential in the Fourier series (\ref{1-12}): in
\cite{bz}, we replace $\nu_{3}(k)\rightarrow \nu_{3}(k)/2^d$,
$\sum^{(2\pi)}\rightarrow \sum^{(\pi)}$ and obtain $E(k)$
(\ref{4-18}) and $E_{0}$ (\ref{1-4b}) \mbox{with $a_{2}(k)$
(\ref{1-41}).}

Relations  $\partial \langle \hat{H}\rangle /\partial
\nu_{3}(k)=\langle \partial \hat{H}/\partial \nu_{3}(k)\rangle $,
$S(k)=$ $=\langle \rho_{\textbf{k}}\rho_{-\textbf{k}} \rangle$ and
(\ref{1-13}) yield \cite{yuv2} $S(k) = 1+(2^{d+1}/n)\,\times$
$\times\,(\partial E_{0}/\partial \nu_{3}(k))$. Using this result,
(\ref{1-33}), and (\ref{1-4b}), we find the connection between
$a_{2}(k)$ and the structural factor $S(k)$.
 In the zero approximation, this connection is the same as for periodic BCs:  $1-2a_{2}(k)\approx$ $\approx 1/S(k)$. With regard
for this relation and (\ref{1-41}), formula (\ref{4-18}) is
transformed into the Feynman formula\vspace*{-1mm}
\begin{equation}\label{4-19}
 E(k)\approx \frac{\hbar^2 k^2}{2mS(k)},
\end{equation}\vspace*{-4mm}

\noindent which agrees approximately with the experiment.

If one particle is in the box, $k$ is restricted  from above by the
value $k^{\rm max}=\sqrt{6mU_{s}}/\hbar$ (see (\ref{1-22})). For
${\rm He}^4$ atoms at $U_{s}=25\,K$, we have $k^{\rm max}\approx
3.6\,\mbox{\AA}^{-1}$. So, the observed break of the dispersion
curve of He II at $k \approx 3.6\,\mbox{\AA}^{-1}$ can be caused, in
principle, by that a phonon has also some $k^{\rm max}$ at a finite
$U_{s}$, rather than by the decay mechanism
\cite{pitaev}.\vspace*{-2mm}

\section{Comparison\\ of the New and Traditional Solutions}

Expansion (\ref{1-6}), (\ref{pot5}) yields the traditional solution.
For the transition to it, it is necessary to replace $1/2^d
\rightarrow 1$, $\sum^{(\pi)}\rightarrow \sum^{(2\pi)}$ in the
equations. The curves in Figs. 1 and 2 are true also for the
traditional solution, but for $U_{b}$ which is less by a factor of
8. For a cyclic system, we have only this traditional solution. But,
for a system with boundaries, we obtain the traditional and new
solutions.

The principal question is as follows: Which of the solutions is
realized in the Nature for systems with boundaries? We found
numerically that, at a weak interaction, $E_{0}$ is noticeably less
for the new solution (usually by $\sim 2^d$ times): in 2D and 3D
cases for all parameters, and in 1D case almost for all ones. We
studied potentials (\ref{1-39}) and $U_{b}\delta(x/a)$ for 1D and
potential (\ref{1-39}) for 2D. For 3D, we took (\ref{1-39})
and\vspace*{-1mm}
 \begin{equation}\label{5-1}
 U_{3}(\textbf{r}) \approx
\left \{\!\! \begin{array}{ll}
 \displaystyle   U_{b}(1-r^{2}/a^{2}),  &    r\leq a , \\[1mm]
 \displaystyle    U_{bd}\left [\!\left
   (\!1-\frac{a}{r_{0}}\!\right )^{\!\!2}-\left (\!\frac{r}{r_{0}}    -1\!\right )^{\!\!2}\!\right ]\!, & \   a\leq r \leq \tilde{b},
  \\[2mm]
    0,  & \ r>\tilde{b},  \end{array} \right.
\end{equation}\vspace*{-5mm}

\noindent where $\tilde{b}= 2r_{0}-a$. The quantity $E_{0}$ was
calculated by formulas (\ref{1-4b}) and (\ref{1-41}). In 2D,
potential (\ref{1-39}) has the Fourier-transform
\begin{equation}\label{5-3}
\nu_{2}(\textbf{k}) =\frac{2\pi a}{k}(U_{b}-U_{bd})J_{1}(ka)+\frac{2\pi b}{k}U_{bd}J_{1}(kb).
\end{equation}
Here, $J_{1} (p)$ is the Bessel function. We consider the weakness
of the interaction as the smallness of the barrier $U_{b}$ and (for
3D) of the concentration $n$. At a nonweak interaction, $E_{0}$ for
the new solution is sometimes less than that for the traditional
one. But the opposite is also possible; this depends on the
approximation, the parameters and, possibly, the error of
calculations.

The solution with larger $E_{0}$ corresponds to the unstable
ordering. Therefore, the new solution is true  at a weak
interaction; but we don't know which solution is true at a strong
interaction. As was mentioned above, the traditional solution for
the systems with boundaries follows from expansion (\ref{1-6}),
(\ref{pot5}), which creates $2^d -1$ fictitious images and makes the
total interatomic potential cyclic. In such system, there is no
symmetry breaking that corresponds to the boundary, besides the
hand-imposed condition $\Psi =0$. Such statement of the problem is
not quite self-consistent, since we are interested in the
topological effect, but the topology is modeled unproperly. This
calls into question the traditional solution. However, this solution
was found \cite{zero-gasGP1} also from the exact expansion
(\ref{1-6}), (\ref{1-12}) in the Gross--Pitaevskii approach.
Therefore, the traditional solution can also exist in our
approach.\vspace*{-2mm}

 \section{Why Can the Boundaries\\ Affect the Bulk Properties?}
To argue why  the boundaries should not affect the bulk properties,
the following general arguments are presented: 1) for large systems,
the volume of the near-boundary region is negligibly small as
compared with the total volume of the system; 2) the dimensional
formulas of the form $E=Nf(S/N,V/N)$ \cite{land5}; 3) the proofs
\cite{ruelle} of the existence of a thermodynamic limit. These are
not simple questions, and they are slightly studied. We will try now
to clarify them. First, the above-obtained effect of boundaries has
the bulk character (see below); in this case, argument (1) is not
valid. We now answer item (3), which will also answer item (2). It
was proved in monograph \cite{ruelle} that if, for a cyclic system
or a system with boundaries, $N$ and $V$ are unboundedly increased
at $N/V=$~const, then the partition function approaches some
limiting value. If we perform the same transition in the
above-obtained formulas for a cyclic or noncyclic system, then the
equations remain invariant, i.e., the limits exist too. Therefore,
our results agree with those in \cite{ruelle}. However, such limit
is almost obvious \cite{ruelle} without calculations. The following
is not obvious: Do the limits for initially cyclic and initially
noncyclic systems coincide?  This was not considered in
\cite{ruelle}. In our approach, these limits turn out
\textit{different}. Another question is as follows: Do the limits
\cite{ruelle} mean the transition to the infinite system? The
passage to limit was performed by Van Hove and by Fisher. In both
cases, the finite system was considered, and the passage to infinity
is realized in the meaning of the continuous limit $N, V \rightarrow
\infty$. For a cyclic system, we can continuously pass to the
infinite system, whereas it is \textit{impossible} for a system with
boundaries: a boundaries cannot disappear at increasing the system
size, because the boundary is a topological property (this can be
referred to a lot of remarkable paradoxes for infinite sets
\cite{kline}). In this case, the transition to the infinite system
assumes a topological jump. Therefore, the Ruelle limit for a system
with boundaries is the passage to an arbitrarily large, but finite
system. Hence, the validity of the transition from a system with
boundaries to the infinite system \textit{was not} proved in
\cite{ruelle}. But the T-limit is used in physics as a way to avoid
the consideration of boundaries, and it means the transition to
namely the infinite system (without boundaries). Such ``strong''
T-limit assumes that the properties of four following different
systems are identical: finite large cyclic, infinite cyclic, finite
large with boundaries, and infinite noncyclic. This strong
assumption is not proved in the general case, and our result
indicates that this assumption is erroneous for some systems.
Systems with different topologies are described by different
(generally speaking) eigenfunctions. As a result, the values of
$E/N$ (e.g.) for them can be different. From the four
above-mentioned systems, the topologies of only two first systems
are identical. Therefore, the transition between them is rightful,
but the transition between any other systems can lead to a jump of
bulk properties. The analysis in the previous sections indicates
that the transition from a finite cyclic system to a finite system
with boundaries leads to a jump by a factor of $2^{-d}$. The jump at
the transition from an infinite cyclic system to the infinite
noncyclic one consists in the neglect of images (see Section 2).
Sometimes, a cyclic infinite system is described without the account
for images, but this is wrong, in our opinion. Thus, arguments
(1)--(3) do not imply that the boundaries cannot affect the bulk
properties. The weak point of the T-limit is the neglect of the jump
due to the topology; eventually, the main source of difficulties is
the passage to infinity, which is contradictory \cite{kline}. Here,
we consider only the finite systems, for which all things can be
properly defined.

We note that the function $f$ in the formula $E=$ $=Nf(S/N,V/N)$ can depend on the shape of
a boundary, at least as $T \rightarrow 0$. Our calculation for a vessel-parallelepiped has shown that $E_{0}$
and the dispersion law are independent of the ratio of the sizes of a vessel.
Most likely, no dependence on the shape of a boundary exists for vessels of any shape.

How can the topology affect the bulk properties of a system? We
indicate two ways. 1) ``The effect of images''. If two particles are
placed on a finite one-dimensional ring, then a particle acts on
another one from two sides (from $2^{d}$ sides in $d$ dimensions).
After we unclose the ring, the action will be only from one side. By
expanding the interatomic potential in the exact series (\ref{1-6}),
(\ref{1-12}), one can show that the account for $2^{d}-1$ images
transfers expansion  (\ref{1-6}), (\ref{1-12}) in the traditional
one (\ref{1-6}), (\ref{pot5}). We think that just this is the reason
for the difference by the factor $2^{d}$ in the dispersion law. 2)
``The effect of modes''.  On a closed 1D string, the standing waves
with $\lambda = L/j$ ($j$ is an integer) are possible; whereas an
unclosed string admits additionally the standing waves with $\lambda
= 2L/j$. In other words, the number of eigenmodes for a
$d$-dimensional system with boundaries is more by $2^{d}$ times,
than that for the same cyclic system. It is known
\cite{bz,chester1967} that the ground-state WF for interacting
bosons $\Psi_{0}\simeq \exp{(\sum_{\textbf{k}}a_{2}(\textbf{k})
\rho_{\textbf{k}}\rho_{-\textbf{k}}/2+...)} $ is similar to the WF
of the system of interacting oscillators. Mechanics implies that the
eigenfrequencies of the system of interacting oscillators differ
from the frequencies of free oscillators. Moreover, the frequencies
of the system of $N$ oscillators differ from those of the system of
$2^{d}N$ oscillators of the same type. Therefore, the energies of
the lowest states of these two systems must be different. We note
that a phonon can be considered as one more oscillator. It is clear
that if we add the same oscillator to the system of $N$ interacting
oscillators and to the system of $2^{d}N$ oscillators, then the
frequencies of these systems shift by different values. This shift
is the phonon energy. Therefore, the frequencies of phonons with the
same $\textbf{k}$ in systems with different topologies can be
different, though it seems strange at the first sight. Our solution
indicates that the influence of modes is strong at a strong
interatomic interaction. This conclusion is natural. For example,
for the ground state of He II, the stronger the interatomic
interaction, the stronger is the interaction between
``oscillators''. The influence of modes is manifested in that
whether $\textbf{k}$ is quantized as $\pi/L$ or as $2\pi/L$.

The physics of condensed systems is determined by collective modes (waves). Therefore,
the boundaries must (or, at least, can) affect the dispersion law and the ground-state energy.
At high temperatures, the physics is determined by atoms. The standing waves
are modulated by a wall and therefore preserve the memory of the wall,
whereas the atoms lose the memory after several collisions with other atoms. Therefore,
at high $T,$ the effect of modes must not hold. As for the influence of ``images'', the answer is not obvious.
We expect that, at high $T,$ the topology has no influence on the bulk properties.

K. Huang \cite{huang} noted that the boundaries must have no effect (most probably),
since the correlation length $\lambda_{c}$ is much less than the size of the system.
But this $\lambda_{c}$ is obtained from the binary correlation function of \textit{all} atoms.
For the condensate subsystem, $\lambda_{c}$ is equal to the size of the system, as
follows from the definition of a condensate and from the experiments with gases in a trap.
Therefore, if the condensate is ``pricked'' at some place, this will be felt by all condensate atoms.
This consideration led the author to the idea of a possibility of the effect of boundaries.
Now, we suppose that the effect of boundaries has a more general character and is possible without condensate.

 The main point consists in that the effect of boundaries is
related to the topology of a system as a whole, rather than to the
influence of a thin layer of near-surface atoms. This effect is a bulk one.

\section{Comparison with Other Models}

The influence of boundaries was studied within several models \cite{lieb1963,gaudin,stringari1996,zaremba1998,cazalilla2004},
but no effect was found. Let us clarify, why.
In Section 6, we have considered two sources of the effect of boundaries: images and modes.

In the exactly solvable 1D model \cite{lieb1963,gaudin}
for bosons with the point interaction,
the cyclic and noncyclic systems are described by the
Hamiltonian
\begin{equation}\label{10-0}
 \hat{H} = -\sum\limits_{j}\frac{\partial^{2}}{(\partial x_{j})^2} + 2c\sum\limits_{i<j}
\delta(|x_{i}-x_{j}|)
\end{equation}
without images $U(L_{x}-|x_{i}-x_{j}|)$.
The account for images means the necessity to explicitly consider the interaction of the first and $N$-th atoms for the cyclic system.
This adds one equation. However, one can show that it is satisfied identically.
The reason is the following. In order to describe the point interaction of the first and $N$-th atoms, we must consider both atoms
on the segment $x>0$  (then the $N$-th atom becomes the first one, and the first atom becomes the second one) or on the segment
$x<L$  (then the first atom becomes the $N$-th one, and the $N$-th atom becomes the $N-1$-th one).
But such interactions have been already included in the equations \cite{lieb1963}.
For nonpoint atoms, the influence of images would be nonzero.

Can the modes affect?  The solution for the WFs of a cyclic system of point particles is
\cite{lieb1963}:
\begin{equation}\label{10-1}
\Psi_{(k)}(x)=\sum\limits_{P}a(P)Pe^{i\sum\limits_{l=1}^{N}k_{P_{l}}x_{l}}\!,
\end{equation}
where $P$ are all permutations of $k_{l}$.
WF (\ref{10-1}) is a superposition of  states of
$N$ free particles. The energy,
\begin{equation}\label{10-2}
 E=k_{1}^{2}+k_{2}^{2}+... +k_{N}^{2},
 \end{equation}
is the same for all these states.
Since the atom interacts only with the neighbors on the left and on the right, the ground-state WF for cyclic
boundaries must have the form
\begin{equation}\label{10-2}
 \Psi_{0}= \sum\limits_{P} F(x_{1,2},x_{2,3},...,x_{N-1,N},x_{N,1}),
\end{equation}
where $P$ are the  permutations of $x_{j}$, and $x_{i,j}=x_{i}\,-$
$-\,x_{j}$. For noncyclic boundaries, $x_{N,1}$  should be replaced
by $x_{N}, x_{1}$ in (\ref{10-2}). WF (\ref{10-1}) can be
rewritten~as
\begin{equation}\label{10-3}
\Psi_{(k)}(x)=\sum\limits_{P}a(P)Pe^{i\sum\limits_{j=1}^{N-1}
p_{j}(k)x_{j,j+1}+ip_{N}(k)x_{N}}\!,
\end{equation}
where $ p_{j}(k)=\sum_{l=1}^{j}k_{l}$. For the GS, $p_{N}=0$
\cite{lieb1963}. Therefore,\vspace*{-3mm}
\begin{equation}\label{10-5}
\Psi_{(k)}(x)=\sum\limits_{P}a(P)Pe^{i\sum\limits_{j=1}^{N-1}
p_{j}(k)x_{j,j+1}+ip_{N}(k)x_{N,1}}\!,
\end{equation}
which is structure (\ref{10-2}). In the presence of boundaries, the ground-state WF is more complicated, but it is constructed on the basis of
(\ref{10-1}). We now have $p_{N}\neq 0$ \cite{gaudin}, and, therefore,
(\ref{10-3}) does not transit in (\ref{10-5}). These ground-state WFs have nonoscillatory structure,
as opposed to $\Psi_{0}$ of a Bose liquid. Therefore, there is no effect of modes.

Thus, the effect of boundaries is absent in the models by Lieb \cite{lieb1963} and
Gaudin \cite{gaudin}  due to the point interaction.

Though  $\Psi_{0}$ for the systems with point
and nonpoint interactions are different, the Bogolyubov approach \cite{bog} does not use WFs
and allows to describe these systems in the same way.
Therefore, the solutions by Lieb \cite{lieb1963} and Gaudin \cite{gaudin} at a weak  interaction
are close to Bogolyubov  ones.
To consider boundaries, the Bogolyubov method \cite{bog} should be modified.

In the Gross--Pitaevskii approach with point potential, the solutions \cite{stringari1996,zaremba1998}
correspond to the Bogolyubov mode, changed due to the inhomogeneity and (for low-lying modes) a geometry.
For a nonpoint interaction, two solutions were found \cite{zero-gasGP1}: the Bogolyubov mode and the new one (\ref{4-18}).

In work \cite{cazalilla2004}, Haldane's harmonic-fluid approach is applied to Bose and Fermi systems,
and the exact Hamiltonian with the
potential $U(\textbf{r}-\textbf{r}^{\prime})$ was replaced by an approximate one
 (Luttinger liquids).
In this Hamiltonian, the interaction singularities are present only in constants,
and the difference character $(\textbf{r}-\textbf{r}^{\prime})$ of the potential
is absent.  However, the effects of images and modes are
characteristic of a nonlocal Hamiltonian with the interaction term of the type $f(\textbf{r}-\textbf{r}^{\prime})$.

\section{Thermodynamics}

 Under periodic BCs, the thermodynamic quantities are calculated from the free energy [42]:
\begin{equation}\label{6-1}
 F = -k_{B}T\int \ln(1+n_{q})\frac{d\textbf{k}}{(2\pi)^{3}}.
\end{equation}
 Under zero BCs,
law (\ref{1-11}) holds. Therefore, the quantity $2\pi$ in (\ref{6-1}) should be replaced by $\pi$.
However, eight states of a standing phonon with different signs of the components of $\textbf{k}$ are equivalent.
Therefore, we must integrate only over the sector $k_{x}, k_{y}, k_{z}>0$.
We arrive at formula (\ref{6-1}) and the known formulas for entropy, heat capacity, {\it etc}.

\section{Experimental Tests}
Which of two solutions corresponds to experiments? It is not easy to give a reply.
For He II, the basic problem consists in that the higher corrections to the
equations are large, but they are omitted.
We can estimate $E_{0}$  (\ref{1-33}), (\ref{1-4b}) for He II with regard
for $A_{1}\approx 0$ and $a_{2}$ (\ref{1-41})
(two-particle approximation). We use $d=3$
and $\sum\limits^{(\pi)}$ for the new solution in (\ref{1-4b}), (\ref{1-41}) and $d=0$
and $\sum\limits^{(2\pi)}$ for the traditional one.  For the new solution,
$E_{0}$ coincides with the experimental value $E_{0}^{\rm exp}=-7.16\,$K for $ U_{b}=64\,$K and $142\,$K for potentials (\ref{1-39})
 and (\ref{5-1}), respectively (for
$a= 2\,\mbox{\AA}$, $b= 4\,\mbox{\AA}$, $r_{0}= 3.5\,\mbox{\AA}$,
$U_{bd}=-9\,$K, and $\bar{R}= 3.58\,\mbox{\AA}$). For the
traditional solution, $E_{0}$ does not coincide with $E_{0}^{\rm
exp}$  at potential (\ref{1-39}) (for all $ U_{b}$) and coincides at
potential (\ref{5-1}), if $ U_{b}=92\,$K. These estimates do not
involve the higher corrections, which can change strongly the
results. For the dispersion law in the first approximation
\cite{sunakawa1970,camp1973,rovenchak2000,mt2}, the traditional
solution corresponds to experiments for $U_{b}\sim 100 - 300\,$K,
and the new solution should correspond to experiments for
$U_{b}\sim 2000\,$K due to the factor $2^{-d}=1/8$.

The modern data on the potential give $U_{b}\sim 10^{6}\,$K
\cite{aziz1997,szalewicz2012}. The large distinction from the
theoretical estimates ($U_{b}\sim 100 - 2000\,$K) is related to the
neglect of higher corrections and to the possibly overestimated value $
10^{6}\,$K. However, the new solution is closer to $10^{6}\,$K. The
calculations by the Monte-Carlo method were performed under cyclic
BCs; for  $U_{b}\sim 10^{6}\,$K, they give the results in
approximate agreement with experiments (see review
\cite{ceperley1995}).

A gases in traps are nonuniform and strongly localized in space.
Under such conditions, the boundaries should probably not affect the
bulk microstructure of a system (see [38] for more details).

We propose several direct tests. According to (\ref{1-4b}) and
(\ref{4-18}), a system with boundaries has the smallest values of
$E_{0}$ and $E(k)$ in 2D at a weak interaction. At the closure of
one of the coordinates, $E_{0}$ and $E(k)$ increase (in this case,
$\sum_{\textbf{k}}=\sum_{k_{1}}^{(\pi)}\sum_{k_{2}}^{(2\pi)}$ in
(\ref{1-4b})), and they increase more at the closure of the second
coordinate. Therefore, the dispersion curve should be different for
a  monoatomic dilute  He II film on a plane surface ($d=2$ in Eqs.
(\ref{1-4b}), (\ref{4-18})), on the lateral surface of a cylinder
($d=1$), and on the surface of a torus ($d=0$). The heat capacity of
these films at the same temperature must be significantly different.
A more interesting effect is also possible. It is necessary to form
a monoatomic film of He II on the surface of a torus and then to cut
the torus (similarly to the breaking of a ring). As a result, some
boundaries must arise (it is important that the film would not flow
through the boundary and would not join itself inside the torus; to
this end, it is possible to use a Cs knife, since Cs is not wetted
by helium). In this case, the system passes from the state with
$d=0$ to that with $d=1$. Therefore, the energies of phonons and of
the ground state are sharply changed, the ensemble of phonons will
be rearranged, and the temperature of the system will jump by a
value comparable with the initial $T$. If we restore the slit torus,
the inverse processes will run. These phase transitions  can be
called topological ones. One can also  compare
    the \textit{surface} phonon-roton dispersion curve for a
    \textit{thick} closed  He II film with that one for an open film.

Our analysis shows that if the film thickness is more than several
atomic layers, the properties of a helium  film on the cylinder
surface must probably be the same as those of helium in the cylinder
bulk.\vspace*{-2mm}

 \section{A Few Comments}

Of interest  is the following  property. The ground state is
described by WF (\ref{1-23}) with the product of $N$ triple sines
with the wave vector $\textbf{k}_{1}$, and the excited state is
presented by the same WF with the factor $\psi_{\textbf{k}}$. The
sines of the GS set a standing wave in the probability field. In the
zero approximation, $\psi_{\textbf{k}}$ is equal to $
\rho_{-\textbf{k}}$ with permutations and ``is convolved'' with the
sines of the GS, by resulting in a superposition of states, whose
structure coincides with that of $\Psi_{0}$ with the difference that
the wave vector for some sines differs from $\textbf{k}_{1}$. We may
roughly say that \textit{the excitation is simply the replacement
of one of the standing waves of the GS by a wave with larger
$\textbf{k}$}. Such a structure of WFs means that the GS is formed
by $N$ identical standing  ``phonons''  with smallest wave vector
$\textbf{k}_{1}$ (for which a half-wave occupies the system).
 In the excited states, a part of
phonons is replaced by phonons with different $\textbf{k}$.
In this case, the meaning of the notion of
``excitation'' is changed, because the GS becomes, in a certain
sense, ``excited''.

With regard for corrections, the harmonics of the GS are not
identical by structure to the harmonics of excited states.
But the GS is the
state with $N$ \textit{identical} interacting ``phonons'', and their structure
may be different from the structure of a single ``ordinary''
phonon.

While being scattered in He II, neutrons create phonons. This process occurs with the conservation
of the momentum. In our solution, phonons are standing waves
without definite momentum. However, the WF of a phonon is a sum of eight traveling waves, each  with  a
definite momentum.
 In a small region of helium, we see a lot of quasiparticles as moving localized wave packets with
a certain momentum.  Neutrons create, apparently, just such quasiparticles.
However, each quasiparticle is a part of the total system of standing waves filling the vessel.

We now return to the method. Under periodic BCs, the collections of
$\rho_{\textbf{k}_{j}}$ (\ref{1-op}) are functionally independent
\cite{yuv1,yuv2}, and $\textbf{k}_{j}$ are quantized by the
$2\pi/L$-law (\ref{1-7}). But, for the new solution,
$\textbf{k}_{j}$ are quantized sometimes by the $\pi/L$-law
(\ref{1-11}); in this case, the collections of
$\rho_{\textbf{k}_{j}}$ (\ref{1-op}) are not functionally
independent. However, this is not a problem. Indeed, let us consider
a function
\begin{equation}\label{d-1}
F(x)=\sum\limits_{j}A_{j}f_{j}(x).
\end{equation}
If the Schr\"{o}dinger equation with $\Psi(x)=F(x)$ is reduced to
the form
\begin{equation}\label{d-1}
\sum\limits_{l}B_{l}(\{A_{j}\})f_{l}(x)=0,
\end{equation}
and we determine coefficients $ A_{j}$ such that $B_{l}=0$ for all $l$,
then it is obvious that $F(x)$ is a solution
irrespective of whether the functions $f_{j}(x)$ are independent and whether they form a full collection.
We have acted just in this manner.
Under a certain structure of the equation, it is easier to find solutions
if $f_{j}(x)$ have some special structure and are dependent. It seems that this is our case.
We have used the expansion in the collections of $\rho_{\textbf{k}_{j}}$ (\ref{1-op}) with $\textbf{k}\sim \pi j/L$.
A part of them, namely  the collections with  $\textbf{k}\sim 2\pi j/L,$
forms the full collection.

 While this work was under a thorough consideration,
we have solved the problem in an approach with independent basic functions \cite{zero-gas1}.
The solutions coincide with those obtained above.

Additional comments can be found in the arXiv version (v.\,5) of
this paper (in Section X and Appendix).

\section{Conclusions}

We have studied the influence of boundaries on the microstructure of
He II and obtained two solutions for the wave functions and the
dispersion law: the traditional and new ones.\,\,They correspond to
two different orderings of the system.\,\,The new solution is
obtained from a more exact expansion of the interatomic
potential.\,\,At a weak interaction, the ground-state energy is less
for the new solution.\,\,Thus, the boundaries affect strongly the
bulk microstructure of the system.\,\,So, the  models with periodic
boundary conditions should be reconsidered.\,\,Their agreement  with
the experiment may be related to the approximate character of the
models and to the human psychology (fitting parameters, the tuning
of a model or interrupting the study, when the agreement with the
experiment appears, rather than when the model becomes
\mbox{exact}).\looseness=1

Here, we recall the Casimir effect \cite{casimir}
caused by the influence of the boundaries, geometry,
and topology on the eigenmodes of vacuum of the system (see review \cite{mostep}).

Because the effect of boundaries is related to the topology, it should be valid
 for all interacting systems at $T\rightarrow 0$.
The effect must be also present in quantum field theory. Since the
Nature has no infinite and point objects, the elementary particles
must have nonzero sizes. In view of this, the Lagrangian should
contain the interaction term of the form
$U(\textbf{r}-\textbf{r}^{\prime})$, which leads to the effect of
boundaries (for the Universe as a whole as well). The further
studies will help us to better understand this
\mbox{phenomenon.}\looseness=1

\vskip3mm
 {\it The author thanks N.~Iorgov, Yu.~Shtanov, K.~Szalewicz, and S.~Yushchenko for  valuable discussions.
He is grateful also to the anonymous referees for useful remarks}.

\rezume{М.Д. Томченко}{%
МІКРОСТРУКТУРА He II ЗА НАЯВНОСТІ ГРАНИЦЬ} {Досліджено
мікроструктуру системи взаємодіючих бозе-частинок за  нульових
граничних умов, і знайдено два можливих впорядкування. Одне
традиційне, та при слабкій взаємодії характеризується законом
дисперсії Боголюбова $E(k) \approx \sqrt{\left (\frac{\hbar^{2}
k^2}{2m}\right )^{2} +
  q n\nu_{3}(k)\frac{\hbar^2 k^2}{m}}$ ($q=1$). А друге -- нове
та характеризується тим самим законом дисперсії, але з $q=2^{-d}$,
де $d$ -- кількість нециклічних координат. При слабкій взаємодії
енергія основного стану менша для нового розв'язку. Границі
впливають на об'ємну мікроструктуру внаслідок відмінності
 топології замкненої та відкритої систем.}


\vspace*{2mm}

\begin{thebibliography}{200}

\bibitem {balescu} R.~Balescu, {\it Equilibrium and Nonequilibrium Statistical Mechanics},
(Wiley, New York, 1975).\vspace*{0.5mm}

\bibitem {lieb1963}  E.H. Lieb and W. Liniger, Phys. Rev. \textbf{130}, 1605 (1963);
E.H. Lieb, Phys. Rev. \textbf{130}, 1616 (1963).\vspace*{0.5mm}

\bibitem {gaudin}  M. Gaudin, Phys. Rev. A \textbf{4}, 386 (1971);
M. Gaudin, {\it La Fonction d'Onde de Bethe} (Masson, Paris, 1983).\vspace*{0.5mm}

\bibitem {stringari1996}  S. Stringari, Phys. Rev. Lett. \textbf{77}, 2360 (1996).\vspace*{0.5mm}

\bibitem {zaremba1998}  E. Zaremba, Phys. Rev. A \textbf{57}, 518 (1998).\vspace*{0.5mm}

\bibitem {cazalilla2004}  M.A. Cazalilla, J. Phys. B: AMOP \textbf{37}, S1 (2004).\vspace*{0.5mm}

\bibitem {bog} N.N. Bogoliubov, J. Phys. USSR \textbf{11}, 23 (1947).\vspace*{0.5mm}

\bibitem {bz}  N.N. Bogoliubov and D.N. Zubarev,
Sov. Phys. JETP \textbf{1}, 83 (1956).\vspace*{0.5mm}

\bibitem {chester1967} L.~Reatto and G.V. Chester, Phys. Rev.  \textbf{155}, 88 (1967).\vspace*{0.5mm}

\bibitem {camp1973} C.E. Campbell,  Phys. Lett. A \textbf{44}, 471 (1973).\vspace*{0.5mm}

\bibitem{feenb} E. Feenberg,  Ann. Phys. \textbf{84}, 128 (1974).\vspace*{0.5mm}

\bibitem {yuv1} I.A. Vakarchuk  and I.R. Yukhnovskii, Theor. Math. Phys. \textbf{40}, 626 (1979).\vspace*{0.5mm}

\bibitem {gfmc1979} P.A.\,Whitlock, D.M.\,Ceperley, G.V.\,Chester, and M.H.\,Ka\-los, Phys. Rev. B \textbf{19}, 5598 (1979).\vspace*{0.5mm}

\bibitem {gfmc1981} M.H. Kalos, M.A. Lee, P.A. Whitlock, and G.V. Chester, Phys. Rev. B \textbf{24}, 115 (1981).\vspace*{0.5mm}

\bibitem {krot1986} E. Krotscheck,  Phys. Rev. B \textbf{33}, 3158 (1986).\vspace*{0.5mm}

\bibitem {krotscheck1988} J.L. Epstein and E. Krotscheck,  Phys. Rev. B \textbf{37}, 1666 (1988).\vspace*{0.5mm}

\bibitem {swf1994a} T. MacFarland, S.A. Vitiello, L. Reatto, G.V. Chester, and M.H. Kalos,
Phys. Rev. B \textbf{50}, 13577 (1994).\vspace*{0.5mm}

\bibitem {swf1994b} L. Reatto, G.L. Masserini, and S.A. Vitiello, Physica B \textbf{197}, 189 (1994).\vspace*{0.5mm}

\bibitem {rovenchak2000} I.O. Vakarchuk, V.V. Babin, and A.A. Rovenchak, J. Phys. Stud. \textbf{4}, 16 (2000).\vspace*{0.5mm}

\bibitem {mt1}  M. Tomchenko, Low Temp. Phys. \textbf{32}, 38 (2006).\vspace*{0.5mm}

\bibitem {rea}  L. Reatto, J. Low Temp. Phys. \textbf{87}, 375 (1992).\vspace*{0.5mm}

\bibitem {ceperley1995}  D.M. Ceperley, Rev. Mod. Phys. \textbf{67}, 279 (1995).\vspace*{0.5mm}

\bibitem {revAJP}   M.D. Tomchenko, arXiv:cond-mat/0904.4434.\vspace*{0.5mm}

\bibitem {yuv2} I.A. Vakarchuk and I.R. Yukhnovskii, Theor. Math. Phys. \textbf{42}, 73 (1980).\vspace*{0.5mm}

\bibitem {mt2}  M.D. Tomchenko, Ukr. J. Phys. \textbf{50}, 720 (2005).\vspace*{0.5mm}

\bibitem {zar1977} E. Zaremba and W. Kohn, Phys. Rev. B \textbf{15}, 1769 (1977).\vspace*{0.5mm}

\bibitem {lev2011}  K.V. Grigorishin and B.I. Lev, Ukr. J. Phys. \textbf{56}, 1182 (2011).\vspace*{0.5mm}

\bibitem {land3} L.D.~Landau and E.M.~Lifshitz, {\it Quantum Mechanics. Non-Re\-lativistic Theory} (Pergamon Press, New York, 1980).\vspace*{0.5mm}

\bibitem {yush}  S.A. Yushchenko, private communication.\vspace*{0.5mm}

\bibitem {bijl} A. Bijl, Physica \textbf{7}, 869 (1940).\vspace*{0.5mm}

\bibitem {gir1965}   M.D. Girardeau, J. Math.  Phys. \textbf{6}, 1083 (1965).\vspace*{0.5mm}

\bibitem {huang} K. Huang, {\it Statistical Mechanics} (Wiley, New York, 1987), Chapter 19.\vspace*{0.5mm}

\bibitem {aziz1997}   A.R. Jansen and R.A. Aziz, J.~Chem. Phys. \textbf{107}, 914 (1997).\vspace*{0.5mm}

\bibitem {szalewicz2012}  W. Cencek,  M. Przybytek, J. Komasa, J.B. Mehl, B.~Je\-zi\-orski, and K. Szalewicz,
J. Chem. Phys. \textbf{136}, 224303 (2012).\vspace*{0.5mm}

\bibitem {fey1}  R.P. Feynman, Phys. Rev. \textbf{94}, 262 (1954).\vspace*{0.5mm}

\bibitem {fc}  R.P. Feynman and M. Cohen, Phys. Rev. \textbf{102}, 1189 (1956).\vspace*{0.5mm}

\bibitem {pitaev}  L.P. Pitaevskii, Sov. Phys. JETP \textbf{9}, 830 (1959).\vspace*{0.5mm}

\bibitem {zero-gasGP1}  M. Tomchenko,   cond-mat/1404.0557. \vspace*{0.5mm}

\bibitem {land5} L.D.~Landau and E.M.~Lifshitz, {\it Statistical Physics},  Part 1
(Pergamon Press, Oxford, 1980).

\bibitem {ruelle} D.~Ruelle, {\it Statistical Mechanics. Rigorous Results}
(Benjamin, New York, 1969).

\bibitem {kline} M.~Kline, {\it Mathematics. The Loss of Certainty}
(Oxford Univ. Press, New York, 1980),
Chapter~9.

\bibitem {xal} I.M. Khalatnikov, {\it An Introduction to the Theory of Superfluidity}
(Perseus, Cambridge, 2000), Chapter 1.

\bibitem {sunakawa1970}  T. Kebukawa, S. Yamasaki, and S. Sunakawa, Progr. Theor. Phys. \textbf{44}, 565 (1970).

\bibitem {zero-gas1}  M. Tomchenko,  arXiv:cond-mat/1204.2149.

\bibitem {casimir}  H.B.G. Casimir, Proc. Kon. Nederl. Akad. Wet. \textbf{51}, 793 (1948).

\bibitem {mostep} V.M. Mostepanenko and N.N. Trunov, Sov. Phys. Usp. \textbf{31}, 965 (1988).\vspace*{-1.5mm}
\begin{flushright}
{\footnotesize Received 16.09.12}
\end{flushright}
\end{thebibliography}
\end{document}